\documentclass[12pt]{article}
\usepackage{amssymb,color,times,epsfig}
\newcommand{\eqref}[1]{(\ref{#1})}
\newcommand{\Remot}{\mathrm{Re\,}}

\newcommand{\be}{\begin{equation}}
\newcommand{\ee}{\end{equation}}
\newcommand{\ba}{\begin{array}}
\newcommand{\ea}{\end{array}}
\newcommand{\beqa}{\begin{eqnarray}}
\newcommand{\eeqa}{\end{eqnarray}}
\newcommand{\beqas}{\begin{eqnarray*}}
\newcommand{\eeqas}{\end{eqnarray*}}
\newcommand{\beqal}{\begin{lefteqnarray}}
\newcommand{\eeqal}{\end{lefteqnarray}}

\begin{document}

\title{Coherent and squeezed states of quantum  Heisenberg algebras}

\author{Nibaldo Alvarez--Moraga  \thanks{email address:
alvarez@dms.umontreal.ca}\\  \small{ Centre de Recherches
Math\'ematiques et d\'epartement de Math\'ematiques et de
Statistique, }  \\ \small{Universit\'e  de Montr\'eal, C.P. 6128,
Succ.~Centre-ville, Montr\'eal (Qu\'ebec), H3C 3J7, Canada} }

\maketitle

\begin{abstract}
Starting from deformed quantum Heisenberg  Lie algebras some
realizations are given in terms of the usual creation and
annihilation operators of the standard harmonic oscillator. Then
the associated algebra eigenstates are computed and give rise to
new classes of deformed coherent and squeezed states. They are
parametrized by deformed algebra parameters and suitable
redefinitions of them as paragrassmann numbers. Some properties of
these deformed states also are analyzed.

\end{abstract}


\baselineskip 0.75cm

\newpage

\section{Introduction}

It is interesting for theoretical and practical reasons to study
coherent and squeezed states associated to the quantum  Hopf
algebras \cite{kn:Dri86,kn:Reshe,kn:Ma}. The Hopf algebra structure
of a quantum algebra provides us with useful technical elements such
as the coproduct, for exemple. In the case of boson quantum
algebras, the special coproduct properties are useful to
characterize multi-particle Hamiltonians \cite{kn:TsoPaJa}. For
example, in the case of the Poincar\'{e} quantum algebra, the
coproduct have been brought to bear to the study the fusion of
phonons \cite{kn:Cele2}. In general, the concept of deformed quantum
Lie algebras found various applications in quantum optics, quantum
field theory, quantum statistical mechanics, supersymmetric quantum
mechanics and some purely mathematical problems. For instance, in
the case of the $su_q (2)$ algebra, it has been found that the $su_q
(2)$ effective Hamiltonians reproduce accurately the physical
properties of the $su(2) \oplus h(2)$ models \cite{kn:BaCiHeRe}. On
the other side, there are some works showing  that quantum algebras
are connected with paragrassmann algebras \cite{kn:Spiri,kn:Fi}.
Paragrassmann algebras are relevant in the studies of  theories that
show the necessity of unusual statistic \cite{kn:Ru}, for instance,
the studies of anyons and topological field theories
\cite{kn:MaWi,kn:AnBa}.

Now, to associate coherent and squeezed states to a quantum
deformed Lie algebra one can use the algebra eigentates (AES)
technique. The AES associated to a real  Lie algebra have been
defined as the set of  eigenstates  of an arbitrary complex linear
combination of generators of the considered algebra
\cite{kn:Brif}. The AES associated to a quantum real deformed Lie
algebra can be defined in a similar way. Indeed, if $A_k (q), \,
k=1,2, \ldots, n$ denote the generators of this deformed algebra
in a given representation, parametrized by the set of deformation
parameters $q,$ then the AES associated to this deformed algebra
will be given by the set of solutions of the eigenvalue equation
\be \sum_{k=1}^n \alpha_k A_k (q) |\psi \rangle = \lambda |\psi
\rangle, \qquad \alpha_k, \lambda \in {\mathbb C}. \ee

The purpose of this work is to compute the AES of the deformed
quantum Heisenberg Lie algebras \cite{kn:HL94}, obtained  by
applying the R-matrix methods \cite{kn:Dri86}, and find new
classes of deformed harmonic oscillator coherent and squeezed
states. We will see that these states will be  new deformations of
the standard coherent and squeezed  states of the harmonic
oscillator system and we will recover them in the limit when the
deformation parameters go to zero.  The approach of AES also gives
us the possibility to construct, starting from a deformed algebra,
some Hamiltonians, of physical systems to which these deformed
coherent and squeezed states are associated, similarly as for
algebras and superalgebras \cite{kn:NaVh1,kn:NaVh3}.

It is important to mention  that the deformed coherent states
obtained by this method differ from the $q$--deformed coherent
states associated to a $q$-deformed oscillator algebra, which is
not a Hopf algebra, constructed by considering either deformed
exponential functions, eigenstates of a given deformed
annihilation operator, a generalization of the usual form of the
standard coherent states, a resolution of the identity technique
or a generalized group theoretical techniques
\cite{kn:Dellinas,kn:Cquesne,kn:BjuPSt}.

The paper is organized as follows. In  section \ref{sec-two}, a
Fock space representation of deformed quantum algebras associated
to the Heisenberg algebra $h(2)$ is given. In section
\ref{sec-three}, we compute the AES associated to these algebras
and  obtain new classes of deformed coherent and squeezed states
that are true deformations of the standard coherent and squeezed
states associated to the harmonic oscillator system. These states
are parametrized by the deformation parameters which will be
considered as real numbers and also as real paragrassmann numbers.
In section \ref{sec-cuatro}, we compute the product of the
dispersions of the position and linear momentum operators of a
particle in these states when the parameters of deformation are
small. We compare them with the corresponding results obtained in
the minimum uncertainty states \cite{kn:NaVh1}.  Some details of
calculations are presented in the Appendices \ref{sec-appa} and
\ref{sec-appb}. We also  give general expressions of these
dispersions, in the case where a non trivial one parameter algebra
deformation family  is concerned, for all values of the
deformation parameter. Finally, we construct a class of
$\eta$--pseudo Hermitian Hamiltonians \cite{kn:AMostafazadeh} to
which a  subset of these deformed states are the associated
coherent states.

\section{Deformed quantum Heisenberg algebras in the Fock representation space}
\label{sec-two}  We are considering in this work, the  deformed
Heisenberg quantum algebras obtained by V. Hussin and A. Lauzon
\cite{kn:HL94}. They have been obtained using  the well-known
$R$--matrix method \cite{kn:Dri86} and are mainly of two types.
The first one is formed  by the generators $A,B,C$ which satisfy
\be \left[A,B\right] = 0, \qquad \left[B,C\right] = - { 2 z \over
p^2 } ( \cosh( pB ) - 1 ), \qquad \left[A,C\right] = {1 \over p}
\sinh (pB) . \label{com-he1} \ee It is denoted by
 $ {\cal U}_{z,p} \, (h(2)),$ where $p$ and $z$ are different from
 zero.

 Let us mention that  the invertible change of basis \beqa {\tilde A} = A, \qquad {\tilde B} = { 2
\over p} \sinh \left( {p B \over 2} \right), \qquad {\tilde C} = {
1 \over \cosh \left( {p B \over 2} \right) } \ C , \label{optilde}
\eeqa leads to the new deformed algebra  $ \ {\tilde {\cal
U}}_{z,0} \, (h(2)):$ \be [{\tilde A} , {\tilde B}] =0, \qquad
[{\tilde B} , {\tilde C}] = - z {\tilde B}^2, \qquad [{\tilde A} ,
{\tilde C}] = {\tilde B}. \label{com-he1tilde} \ee This means that
we get the same commutation relations as in \eqref{com-he1}
 when $p$ goes to zero. As it has been pointed out by Ballesteros et al. \cite{kn:BaHePr}, here the
$p$ parameter is superfluous and the families of bialgebras $
{\cal U}_{z,p} \, (h(2))$ and  $ {\cal U}_{z,0} \, (h(2))$ are
isomorphic (these families are identified there as of type $I_+ $)
on the condition that the coproduct form stands  invariant
\cite{kn:BaCeOl}.

The second quantum deformation of $h(2)$ is given by  \be
\label{altype2} \left[A,B\right] = \left[B, C\right] =0, \qquad
\left[A,C\right] = {e^{pB} - e^{- qB} \over p+q} .
 \ee and is denoted by ${\cal U}_{p,q} \, (h(2)),$ where
$p,q \ne 0.$ It corresponds to so-called so called type $II$
bialgebras in \cite{kn:BaHePr}. When $p=q,$  we find the quantum
Heisenberg algebra obtained in Celeghini  et al. \cite{kn:Cele}
(see also \cite{kn:BaCeOl}), i.e.,  \be \left[A,B\right] =
\left[B,C\right] =0, \qquad \left[A,C\right] = {1 \over p} \sinh
(p B). \label{cel-et-al} \ee

  Let us now give a boson realization of these
deformed Lie algebras, in terms of the usual creation operator,
$a^\dagger, $ and annihilation operator, $a,$ associated to the
standard quantum harmonic oscillator system. For ${\tilde {\cal
U}}_{z,0} \, (h(2))$ given in \eqref{com-he1tilde}, it is given by
\be \label{cas1} \tilde A = - a^\dagger, \qquad \tilde B = e^{z
a^\dagger}, \qquad \tilde C = e^{z a^\dagger} \ a . \ee

From \eqref{optilde} and  \eqref{cas1},  we thus get  a
realization of ${\cal U}_{z,p} \, (h(2))$ as \be
\label{op-def-one-zp} A = - a^\dagger, \qquad B = {2\over p}
\sinh^{-1} \left( {p \over 2} e^{z a^\dagger} \right), \qquad
  C = e^{ z a^\dagger} \sqrt{1+ {\left({p \over 2} e^{z
a^\dagger}\right)}^2 } a. \ee

Another realization of ${\tilde {\cal U}}_{z,0} \, (h(2))$  is \be
\label{cas2} \tilde A = a, \qquad \tilde B = e^{- z a}, \qquad
\tilde C = a^\dagger \ e^{-za}. \ee We thus get another
realization of ${\cal U}_{z,p} \, (h(2))$ as \be
\label{op-def-two-zp} A = a, \qquad \qquad B = {2\over p}
\sinh^{-1} \left( {p \over 2} e^{- z a} \right), \qquad C =
a^\dagger e^{- z a} \sqrt{1+ {\left({p \over 2} e^{-z a}\right)}^2
}.  \ee

When $z$ goes to zero, the operators \eqref{op-def-one-zp} become
\be \label{def1} A=- a^\dagger , \qquad B= {2 \over p} \sinh^{-1}
\left({p\over 2} \right) I, \qquad C = \sqrt{1+ {p^2 \over 4}} a,
\ee while the operators \eqref{op-def-two-zp} become \be
\label{def2} A = a, \qquad B= {2 \over p} \sinh^{-1} \left({p\over
2} \right) I, \qquad C = \sqrt{1+ {p^2 \over 4}} a^\dagger. \ee
The operators  \eqref{def1} or \eqref{def2} thus  constitute a
realization of deformed Heisenberg algebra \eqref{cel-et-al}. When
$p$  goes to zero, we regain $h(2).$

The algebra  \eqref{altype2} is clearly  isomorphic to $h(2)$ if
we introduce \be \tilde A =A, \qquad \tilde C = C, \qquad \tilde B
= {e^{pB} - e^{-qB} \over p+q}. \ee

So to obtain  new class of deformed  coherent and squeezed states
using the AES method we will deal in the following with ${\tilde
{\cal U}}_{z,0} \, (h(2))$ and ${\cal U}_{z,p} \, (h(2)).$

\section{AES and deformed coherent and squeezed states}
\label{sec-three} In this section, we compute the AES associated
to ${\tilde {\cal U}}_{z,0} \, (h(2))$ and ${\cal U}_{z,p} \,
(h(2)), $ using the representations obtained in the preceding
section. We thus get new classes of deformed coherent and squeezed
states associated to the harmonic oscillator system.

\subsection{Deformed algebra eigenstates for \mathversion{bold} ${\tilde {\cal U}}_{z,0} \,(
h(2))$} \label{sec-aes-He} We start with ${\tilde {\cal U}}_{z,0}
\,( h(2))$ as given by \eqref{com-he1tilde} using the realizations
\eqref{cas1} and \eqref{cas2}. The AES are thus defined as the set
of solutions of the eigenvalue equation \be \label{aes-10}
[\alpha_+ {\tilde A} + \alpha_0 {\tilde B} + \alpha_-  {\tilde C}
] |\psi\rangle = \alpha |\psi\rangle, \qquad \alpha_- , \alpha_0 ,
\alpha_+ , \alpha \in {\mathbb C}. \ee

\subsubsection{Deformed harmonic oscillator coherent and squeezed
states}\label{sub-sec-coh-squee} Let us take first the
realization \eqref{cas1}. Thus, if $\alpha_- \ne 0, $ equation
\eqref{aes-10} can be written in the form \be \label{eigen-10}[
e^{z a^\dagger} a + \mu a^\dagger + \nu e^{z a^\dagger}] |\psi
\rangle = \lambda |\psi \rangle, \qquad \mu,\nu, \lambda \,  \in
{\mathbb C}. \ee By defining \be \label{psi-varphi} |\psi\rangle =
e^{- \nu a^\dagger} |\varphi\rangle  \ee and  using $ e^{-\nu
a^\dagger} \, a \,e^{\nu a^\dagger}= a + \nu, $ equation
\eqref{eigen-10} can be reduced to \be \label{reduce-001}[ e^{z
a^\dagger} a + \mu a^\dagger ] |\varphi \rangle = \lambda |\varphi
\rangle, \qquad \mu, \lambda \, \in {\mathbb C}. \ee To solve this
eigenvalue equation, let us consider the Bargmann space ${\cal F}$
of analytic functions $f(\xi)$ ($ \xi \, \in {\mathbb C}),$
provided with the scalar product \be (f_1, f_2) = \int_{{\mathbb
C}} \overline{f_1 (\xi) } f_2 (\xi) e^{-{\bar \xi} \xi} {d {\bar
\xi} d\xi \over 2 \pi i}, \qquad \forall \, f_1,f_2 {\in \cal F}.
\ee It is well-know that any function $f \in {\cal F}$ can be
expressed as a linear combination of orthonormalized  functions
$u_n (\xi)= {\xi^n \over \sqrt{n!}}, \, n=0,1,2, \ldots,$
verifying \be \label{ortho-umn} (u_m, u_n) = \int_{{\mathbb C}}
\overline{u_m (\xi)} u_n (\xi) e^{-{\bar \xi} \xi} {d {\bar \xi}
d\xi  \over 2 \pi i} = \delta_{mn}, \ee that is \be f(\xi) =
\sum_{n=0}^\infty c_n u_{n} (\xi), \ee with \be c_n =
\int_{{\mathbb C}} \overline{u_n (\xi)} f (\xi) e^{-{\bar \xi}
\xi} {d {\bar \xi} d\xi  \over 2 \pi i}. \ee

Let us assume a solution of \eqref{reduce-001} of the type \be
\label{type-sol} | \varphi \rangle = \sum_{n=0}^{\infty} c_n
|n\rangle, \ee where the set of states $\{ |n\rangle
\}_{n=0}^{\infty}$ form the basis of the standard Fock oscillator
space, verifying the orthogonality relation \be \langle m
|n\rangle =\delta_{mn}. \label{st-mn-ortho} \ee As usually, the
action of the operators $a$ and $a^\dagger $ on these states is
given by \be a | n \rangle = \sqrt{n} |n-1\rangle, \qquad
a^\dagger | n \rangle = \sqrt{n+1} |n+1\rangle.  \ee Let us take
$| {\bar \xi} \rangle $ to be the standard coherent states
associated to the harmonic oscillator system, that is \be | {\bar
\xi} \rangle = e^{{\bar \xi} a^\dagger} |0 \rangle =
\sum_{n=0}^{\infty} {{(\bar \xi )}^n \over \sqrt{n!}} |n\rangle.
\ee Then, according to the orthogonality property
\eqref{st-mn-ortho}, the projection of $|\varphi \rangle$ on the
coherent state $| {\bar \xi} \rangle $ is given by the analytic
function \be \label{var-xi} \varphi (\xi) = \langle {\bar \xi} |
\varphi \rangle =\sum_{n=0}^{\infty} c_n u_{n} (\xi). \ee The
action of the operators $a^\dagger $ and $a$ in this
representation corresponds to  \be \label{act-xi}\langle {\bar
\xi} | a^\dagger |\varphi \rangle = \xi \varphi (\xi), \qquad \;
\langle {\bar \xi} | a |\varphi \rangle= {d \varphi \over d\xi}
(\xi).  \ee respectively. Thus, by projecting both sides of the
eigenvalue equation \eqref{reduce-001} on the coherent states $|
{\bar \xi} \rangle$ and then using \eqref{act-xi}, we can write it
as  \be \label{eigen-fock} \left( e^{z\xi} {d\over d\xi} + \mu
\xi\right) \varphi (\xi) = \lambda \varphi (\xi). \ee The general
solution of this differential equation is given by \be \varphi
(\xi) = C_0 (\lambda,\mu, z ) \, \exp\left(\sum_{k=0}^{\infty}
{{(-z\xi)}^k \over (k+1)!} \left( \lambda \xi - {k+1 \over k+2}\mu
\xi^2 \right) \right), \label{solgen-varphi} \ee where $ C_0 $ is
an arbitrary constant which can be fixed from the normalization
condition \be \label{con-nor-var}( \varphi , \varphi) =
\int_{{\mathbb C}} \overline{\varphi (\xi )} \varphi (\xi)
e^{-{\bar \xi} \xi} {d {\bar \xi} d\xi  \over 2 \pi i} =1.\ee Let
us notice that in the particular limit when $z$ goes to zero, the
solution \eqref{solgen-varphi}, becomes the symbol for the
squeezed states \cite{kn:Dodo} associated to the standard harmonic
oscillator, that is \be \label{sym-squee} \varphi (\xi) = C_0
(\lambda,\mu,0) \, \exp\left( \lambda \xi - {\mu \over 2} \xi^2
\right) . \ee This quantity is normalizable only if $|\mu| <1 $
\cite{kn:NORu}.

When $z\ne 0,$ the solution \eqref{solgen-varphi} can be written
in the form \be \varphi (\xi) = C_0 (\lambda, \mu, z ) \,
\exp\left( {\lambda \over z} - {\mu \over z^2} \right) \,
\exp\left( e^{-z \xi} {(\mu - \lambda z + \mu z \xi) \over z^2}
\right). \label{solgen-varphi2} \ee

Going back to the expression \eqref{var-xi}, we get the
coefficients $c_n, \, n=0,1,\ldots, $ as \beqa c_n =
\int_{{\mathbb C}} \overline{u_n (\xi )} \varphi (\xi)
 e^{-{\bar \xi} \xi} {d {\bar \xi} d\xi  \over
2 \pi i}  &=& C_0 (\lambda, \mu, z ) \, \exp\left( {\lambda \over
z} - {\mu \over z^2} \right)  \nonumber  \\ & & \int_{{\mathbb C}}
{{\bar \xi}^n \over \sqrt{n!}} \exp\left( e^{-z \xi} {(\mu -
\lambda z + \mu z \xi) \over z^2} \right)
 e^{-{\bar \xi} \xi} {d {\bar \xi} d\xi \over
2 \pi i} .  \eeqa By using the polar change of variables  $\xi =
\rho e^{i \vartheta},$  this last equation can be written in the
form \beqa c_n &=&  C_0 (\lambda, \mu, z ) \,  \exp\left( {\lambda
\over z} - {\mu \over z^2} \right)  \nonumber \\ & &
\int_{0}^\infty \int_{0}^{2 \pi} { \rho^{n+1} e^{- \rho^2 } \over
\sqrt{n!}} e^{-i n \vartheta} \exp\left( {e^{-z \rho e^{i
\vartheta}} \over z^2} ( \mu - \lambda z + \mu z \rho e^{i
\vartheta})\right) { d\rho d\vartheta \over \pi} . \eeqa Let us
write the exponential factor in the form \beqa && \nonumber
 \exp\left( {e^{-z \rho e^{i \vartheta}} \over z^2} ( \mu -
\lambda z + \mu z \rho e^{i \vartheta})\right)  \\ \nonumber &=&
\sum_{k=0}^{\infty} { \exp\left(-z k \rho e^{i \vartheta}\right)
\over k! } {\left( \mu - \lambda z + u z \rho e^{i \vartheta}
\over z^2\right) }^{k} \nonumber \\   &=& \sum_{k,l=0}^{\infty}
\sum_{m=0}^{k} {k \choose m}  \rho^{l+m} e^{i (l+m) \vartheta} {{(
- z k )}^l  {(\mu z)}^m {(\mu - \lambda z)}^{k-m}\over k! \, l! \,
z^{2k}} \eeqa to get \beqa c_n &=&
 C_0 (\lambda, \mu, z ) \,  \exp\left( {\lambda
\over z} - {\mu \over z^2} \right) \,  \sum_{k,l=0}^{\infty}
\sum_{m=0}^{k} {k \choose m} {{( - z k )}^l {(\mu z)}^m {(\mu -
\lambda z)}^{k-m}\over \sqrt{n!} \; k! \, l! \, z^{2k}} \nonumber
\\ && \left(\int_{0}^{\infty} \rho^{m+l+n+1} e^{- \rho^2}
d\rho\right) \left(\int_{0}^{2 \pi} e^{i (l+m-n)\vartheta}
{d\vartheta \over \pi} \right). \eeqa Using the known results \be
\int_{0}^{2 \pi} e^{i (l+m-n)\vartheta} {d\vartheta \over \pi} = 2
\delta_{l+m-n,0}, \ee \be\int_{0}^{\infty} \rho^{m+l+n+1} e^{-
\rho^2} d\rho = {1\over 2} \Gamma\left({m+l+n \over 2} +1
\right),\ee and performing the sum over the index $l,$ the
expression for the coefficients $c_n$ reduces to  \be
\label{coeff-cn} c_n =   C_0 (\lambda, \mu, z ) \,  \exp\left(
{\lambda \over z} - {\mu \over z^2} \right) {z^n \over \sqrt{n!}}
\, \sum_{k=0}^{\infty} \sum_{m=0}^{k_{<}} {n \choose m} {{( -k
)}^{n-m} \over (k-m)!} {\left({\mu \over z^2}\right)}^m
{\left({\mu\over z^2} - {\lambda \over z} \right)}^{k-m} ,\ee
where $k_{<}$ denotes the minimum between $k$ and $n.$  This last
expression can be written in the form \be c_n =   C_0 (\lambda,
\mu, z ) \, {z^n \over \sqrt{n!}} \, \sum_{m=0}^{n}
\sum_{j=0}^{n-m}  {n \choose m} {(-1)}^{n-m} \upsilon_{mj}
{\left({\mu \over z^2}\right)}^m    {\left({\mu\over z^2} -
{\lambda \over z} \right)}^j , \label{mejorcn} \ee where the
coefficients $\upsilon_{mj}$ are obtained from  \be {k^{n-m} \over
(k-m)! } = \sum_{j=0}^{n-m} {\upsilon_{mj} \over (k-m-j)!}. \ee

Thus the coefficients $c_n, $ $n=1,2,\ldots, $ represent
polynomials of degree $n-1$ in the z variable. For example,  $ c_1
= \lambda C_0, $ \be c_2 = C_0 \sqrt{2!} \left[ \left({\lambda^2
\over 2!} - {\mu \over 2}\right) - {\lambda \over 2 } z \right],
\quad \nonumber c_3 = C_0 \sqrt{3!} \left[ \left( {\lambda^3 \over
3!} - {\mu \lambda \over 2} \right) + \left({\mu \over 3} -
{\lambda^2 \over 2}\right) z + {\lambda \over 6} z^2 \right]. \ee

The normalization constant $C_0$ can be now computed. Indeed,
inserting \eqref{mejorcn} into \eqref{var-xi} and the resulting
expression into the normalization condition \eqref{con-nor-var},
using the orthogonality relation \eqref{ortho-umn}, we get \beqa
\nonumber C_0 \, ( \lambda, \mu, z ) &=& \Biggl[
\sum_{n=0}^{\infty} \, {z^{2n} \over n!} \, \sum_{m=0}^{n}
\sum_{r=0}^{n} \sum_{j=0}^{n-m} \sum_{l=0}^{n-r} {n \choose m}{n
\choose r} {(-1)}^{m+r} \upsilon_{mj} \upsilon_{rl} \\ & & {\left(
{\mu \over z^2 } \right)}^m {\left( {{\bar \mu} \over z^2 }
\right)}^r {\left({\mu\over z^2} - {\lambda \over z} \right)}^j
{\left({{\bar \mu} \over z^2} - {{\bar \lambda} \over z}
\right)}^l \Biggr]^{-{1 \over 2 } } ,
 \label{nor-fac}\eeqa  which has been chosen real. The
 convergence of these series it not easy to determine.
 In the case where $z=0,$ as we have already mentioned, the series $\sum_{n=0}^\infty {|c_n|}^2$ converges for all
 $\lambda$ provided  that $|\mu| < 1.$ In the case $\mu=0,$  this series becomes
 \be
\sum_{n=0}^\infty {|c_n|}^2 = {|C_0 (\lambda,z)|}^2
\exp\left({\lambda \over z}\right) \sum_{n=0}^\infty {{\biggl(- {
\lambda \over z}\biggr)}^{n} \over
 n!} \exp\left( -{{\bar \lambda} \over z} \sum_{k=1}^\infty {{(z^2
 n)}^k \over k!}\right).
 \ee It converges for all $z > 0$ provided that the phase $\theta$ in
$\lambda= \beta e^{i \theta}$ satisfies $ - {\pi \over 2} \le
\theta \le {\pi \over 2}, $ whereas for all  $z<0, $ it converges
if $ {\pi \over 2} \le \theta \le {3 \pi \over 2}. $

Finally, we can show that the normalized algebra eigenstates
$|\varphi \rangle,$ solving \eqref{reduce-001} , can be expressed
in terms of a deformed squeezed operator acting on  the ground
state of the standard harmonic oscillator, that is \be |\varphi
\rangle =  C_0 \, ( \lambda, \mu, z )\exp \left(
\sum_{k=0}^{\infty} {{(-z a^\dagger )}^k \over (k+1)!} \left(
\lambda a^\dagger - { k + 1 \over k+2 } \mu {(a^\dagger)}^2
\right) \right) |0 \rangle. \label{re-norm-squee} \ee Also,
combining this last equation with equation \eqref{psi-varphi}, we
get the algebra eigenstates solving \eqref{eigen-10} to be the
deformed coherent states \be |\psi \rangle =  N_0  ( \lambda, \mu,
\nu,  z ) \,\exp \left( \sum_{k=0}^{\infty} {{(-z a^\dagger )}^k
\over (k+1)!} \left( \lambda a^\dagger - { k + 1 \over k+2 } \mu
{(a^\dagger)}^2 \right) \right) e^{-\nu a^\dagger}|0 \rangle,
\label{eigen-10-aes} \ee where $N_0 \, ( \lambda, \mu, \nu, z )$
is a normalization constant which can computed in the same way as
$ C_0 \, ( \lambda, \mu, z ).$

\subsubsection{Perturbed squeezed states}
\label{sec-perturba-z-real} Let us now  assume that $z$ is a
small perturbation parameter of order $k_0 -1$, where $k_0$ is an
integer greater or equal to $2$. From \eqref{re-norm-squee},
neglecting the terms containing the power of $ z $ greater than
$k_0 -1$,  we can write \beqa \nonumber |\varphi \rangle &
\thickapprox &  C_0 (\lambda, \mu, z, k_0) \Biggl[ 1 +
\sum_{k=1}^{k_0 - 1} {{(-z a^\dagger )}^k \over (k+1)!} \left(
\lambda a^\dagger - { k + 1 \over k+2 } \mu {(a^\dagger)}^2
\right) \\ &+& \cdots + { 1 \over (k_0 -1)!} {\Biggl( {-z
a^\dagger \over 2!} \left(\lambda a^\dagger - {2\over 3} \mu
{(a^\dagger)}^2 \right) \Biggr)}^{k_0 -1} \Biggr] \exp\left(
\lambda a^\dagger - {\mu \over 2} {(a^\dagger)}^2 \right) |0
\rangle. \label{def-squee-st}\eeqa

These states can be normalized in the standard  form. For
instance, when $k_0 = 2, $ $\mu = \delta e^{i \phi},$ $\lambda =
\beta e^{i \theta}, $ where $\phi$ and $\theta $ are real phases,
$ 0 \le \delta <  1, $ and $\beta \ge 0,$ a normalized version of
the deformed squeezed states \eqref{def-squee-st}, is given by
\beqa \nonumber |\varphi \rangle &\thickapprox& \Omega (\delta,
\phi, \beta, \theta) \left[ 1 + z \left( {\delta e^{i \phi} \over
3} {(a^\dagger)}^3 - {\beta e^{i \theta} \over 2} {(a^\dagger)}^2
\right)  \right] \\ & & S \left( - \arctan (\delta )e^{i \phi}
\right) D \left( {\beta e^{i\theta} \over \sqrt{1 - \delta^2}}
\right) |0 \rangle, \label{nor-z-def} \eeqa where \beqa \Omega
(\delta, \phi, \beta, \theta) &=& 1  + {z \beta \over 2
{(1-\delta^2)}^2} \Biggl[ \left( 2 \delta^2 +
\beta^2 \left( {1 + \delta^2 \over  1-\delta^2} \right)  \right)  \cos\theta  \nonumber \\
&-&  \delta \left( 1 + \delta^2 +
 { 2 \beta^2 \over 1-\delta^2 } \right) \cos(\phi - \theta)
\nonumber \\ &+& \delta^2 \beta^2 \left( 1  + {2 \delta^2 \over
3(1-\delta^2)} \right) \cos(2 \phi - 3 \theta) - {2\delta \beta^2
\over 3(1-\delta^2)} \cos(\phi - 3 \theta) \Biggr]. \eeqa Here $
S(\chi) = \exp\left[ - \left(  \chi {{(a^\dagger)}^2 \over 2} -
{\bar \chi} {a^2 \over 2} \right)\right]  $ is the standard
unitary squeezed operator \cite{kn:ruso}  and  $ D (\lambda) =
\exp\left( \lambda a^\dagger - {\bar \lambda} a \right) $  the
standard displacement operator \cite{kn:pere}.

\subsubsection{Deformed squeezed and coherent states parametrized by paragrassmann numbers}
\label{sec-paragra} Let us now use the  realization \eqref{cas2}
of ${\tilde {\cal U}}_{z,0} \, (h(2))$. In the case $\alpha_+ \ne
0, $ equation \eqref{aes-10} can be now written in the form \be
\label{eigen-100}[a + \mu a^\dagger e^{- z a} + \nu e^{-z a}]
|\psi \rangle = \lambda |\psi \rangle, \qquad \mu,\nu, \lambda \,
\in {\mathbb C}. \ee There are two types of equations to solve.
The first type is obtained when  $\mu \ne 0 $ and $\nu \ne 0.$ We
can take \be |\psi \rangle = \exp\left({\nu \over \mu} a \right)
|\varphi\rangle \ee and use the relation, $ \exp\left(- {\nu \over
\mu} a \right) a^\dagger \exp\left({\nu \over \mu} a \right)  =
a^\dagger - {\nu \over \mu}, $ to reduce \eqref{eigen-100} to the
form \be \label{eigen-200} [a + \mu a^\dagger e^{- z a} ] |\varphi
\rangle = \lambda |\varphi \rangle, \qquad \mu, \lambda \, \in
{\mathbb C}. \ee If $\nu =0$ and $\mu \ne 0,$ we see from
\eqref{eigen-100} that the same type of eigenvalue equation must
be solved. The second type is obtained when $\mu =0. $ The
eigenvalue equation is \be \label{eigen-30} [a + \nu e^{- z a} ]
|\psi \rangle = \lambda |\psi \rangle, \qquad \nu, \lambda \, \in
{\mathbb C}.\ee

We begin with the resolution of Equation \eqref{eigen-200}. Let us
assume $| \varphi \rangle $ to be again a solution of the type
\eqref{type-sol}. Thus, proceeding as in the preceding section,
the eigenvalue equation satisfied by the symbol $\varphi (\xi),$
in the Bargmann representation, is given by \be \label{fock-para}
 \left({d \over d \xi} + \mu  \, \xi \,  e^{- z {d \over d \xi}} \right) \varphi (\xi)
 = \lambda \varphi (\xi ), \qquad \mu, \lambda \, \in {\mathbb
C}.  \ee  To solve this equation, let us  assume that $z$ is a
real paragrassmann number \cite{kn:Fi,kn:Ru}, that is $z^{k_0}=0,$
for some integer $k_0 \ge 1.$ A detailed procedure of resolution
of this equation is given in the Appendix \ref{sec-appa}. Let us
notice that the case $k_0=1,$ i.e., $z=0,$ is somewhat trivial
since the eigenfunctions $\varphi (\xi)$ solving
\eqref{fock-para}, are given by the standard squeezed symbol
\eqref{sym-squee}. When $k_0=2,$ or $z^2=0,$ i.e., when $z$ is a
odd Grassmann number \cite{kn:Dewit,kn:Corn}, the eigenvalue
equation \eqref{fock-para} becomes \be \left( (1- \mu z \xi) {d
\over d \xi} + \mu  \, \xi \right) \varphi (\xi)
 = \lambda \varphi (\xi ), \qquad \mu, \lambda \in {\mathbb
C}. \ee There are two independent solutions (see Appendix
\ref{sec-appa}).  The normalizable solution of this eigenvalue
equation,  is given by the deformed squeezed symbol \be \varphi
(\lambda, \mu, z) (\xi) = C_0 (\lambda,\mu,z) \left[1 +  z \mu
\left( \lambda {\xi^2 \over 2} - \mu {\xi^3 \over 3}
\right)\right] \exp\left(\lambda \xi - {\mu \over 2} \xi^2
\right). \label{nor-sol-uno}\ee A normalized version of these
states, in the Fock space representation, is given by \beqa
\nonumber |\varphi \rangle &=& {\tilde \Omega} (\delta, \phi,
\beta, \theta) \left[1 + z \delta \left( {\delta e^{2i \phi} \over
3} {(a^\dagger)}^3 -  {\beta e^{i ( \theta + \phi )} \over 2}
{(a^\dagger)}^2 \right)
 \right] \\ & & S \left( - \arctan (\delta )e^{i \phi} \right) D \left(
{\beta e^{i\theta} \over \sqrt{1 - \delta^2}} \right) |0 \rangle,
\eeqa where $\lambda$ and $\mu$ have been chosen as in the
preceding subsection  and \beqa {\tilde \Omega} (\delta, \phi,
\beta, \theta) &=& 1 - {z \delta \beta \over 2 {(1-\delta^2)}^2}
\Biggl[ \left( 2 \delta^2 +
\beta^2 \left( {1 + \delta^2 \over  1-\delta^2} \right)  \right)  \cos(\theta- \phi)  \nonumber \\
&-&  \delta \left( 1 + \delta^2 +
 { 2 \beta^2 \over 1-\delta^2 } \right) \cos \theta
\nonumber \\ &+& \delta^2 \beta^2 \left( 1  + {2 \delta^2 \over
3(1-\delta^2)} \right) \cos(\phi - 3 \theta) - {2\delta \beta^2
\over 3(1-\delta^2)} \cos(2 \phi - 3 \theta) \Biggr].\eeqa When
$k_0 = 3, $ or $z^3=0,$ the eigenvalue equation \eqref{fock-para}
becomes the second order differential equation \be
\label{sec-order-eq}\left( {1\over 2} \mu z^2 \xi {d^2 \over
d\xi^2} + (1- \mu z \xi) {d \over d \xi} \right) \varphi (\xi)
 = (\lambda -   \mu \, \xi ) \varphi (\xi ), \qquad \mu, \lambda, \, \in {\mathbb
C}. \ee According to the results obtained in Appendix
\ref{sec-appa}, the general solution of this equation can be
expanded in the form \be \label{sol-type-3} \varphi (\xi) =
\varphi_0 (\xi) + z \varphi_1 (\xi) + z^2 \varphi_2 (\xi), \ee
with \beqa \varphi_0 (\xi) &=& C_0 \exp\left(\lambda \xi - \mu
{\xi^2 \over
2}\right) , \\
\varphi_1 (\xi) &=& \left[ \mu \left( \lambda {\xi^2 \over 2} -
\mu {\xi^3 \over 3} \right) C_0 + C_1 \right] \exp\left(\lambda
\xi - \mu {\xi^2 \over 2}\right) , \\ \nonumber \varphi_2 (\xi)
&=& \Biggl[ \left( \mu (\mu - \lambda^2 ) {\xi^2 \over 4} +
{2\over 3} \mu^2 \lambda \xi^3 +  \mu^2 (\lambda^2 - 3 \mu )
{\xi^4 \over 8} - \lambda \mu^3 {\xi^5 \over 6} +
\mu^4  {\xi^6 \over 18}  \right) C_0 \\
&+& \mu \left( \lambda {\xi^2 \over 2} - \mu {\xi^3 \over 3} \right)
C_1 + C_2 \Biggr] \exp\left(\lambda \xi - \mu {\xi^2 \over
2}\right), \eeqa  where $C_0, $ $C_1$ and $C_2$ are arbitrary
integration constants. Three independent solutions may thus be
obtained. The first one is obtained by taking $C_1 = C_2 =0.$ We get
\beqa \nonumber   \varphi (\xi) &=& C_0  \Biggl[ 1 + z \mu \left(
\lambda {\xi^2 \over 2} - \mu {\xi^3 \over 3} \right) + z^2 \Biggl(
\mu (\mu -
\lambda^2 ) {\xi^2 \over 4} + {2\over 3} \mu^2 \lambda \xi^3 \\
\nonumber &+&  \mu^2 (\lambda^2 - 3 \mu ) {\xi^4 \over 8} -
\lambda \mu^3 {\xi^5 \over 6} + \mu^4  {\xi^6 \over 18} \Biggr)
\Biggr] \exp\left(\lambda \xi - \mu {\xi^2 \over 2}\right)\\ &=&
C_0 \exp\left[ z \mu \left( \lambda {\xi^2 \over 2} - \mu {\xi^3
\over 3} \right) + z^2 f(\xi)\right]\exp\left(\lambda \xi - \mu
{\xi^2 \over 2}\right),
 \eeqa
where \be f(\xi) = \Biggl( \mu (\mu - \lambda^2 ) {\xi^2 \over 4}
+ {2\over 3} \mu^2 \lambda \xi^3  - 3 \mu^3 {\xi^4 \over 8}
\Biggr). \ee This solution can be  normalized and represents a
second order paragrassmann deformation of squeezed states
associated to the standard harmonic oscillator.

The other independent solutions  are given respectively by \be
\varphi (\xi) = C_1 \, z \Biggl[ 1 + z \mu \left( \lambda {\xi^2
\over 2} - \mu {\xi^3 \over 3} \right) \Biggr]   \exp\left(\lambda
\xi - \mu {\xi^2 \over 2}\right) \ee and \be \varphi (\xi) = C_2
\,  z^2
 \exp\left(\lambda \xi - \mu {\xi^2 \over 2}\right). \ee These
 solutions can not be normalized since $z^k, \, k=1,2, $ are not invertible paragrassmann numbers and $ z^k = 0, \,
k= 3,4, \ldots. $

The higher order  paragrassmann deformations of the squeezed
states associated to the standard harmonic oscillator can be
obtained following a similar procedure (see Appendix
\ref{sec-appa}).

In the case of eigenvalue equation \eqref{eigen-30}, the
differential equation to solve is given by \be
\label{eigen-30-diff} \left( {d\over d\xi} + \nu e^{- z {d\over
d\xi}} \right)
 \varphi (\xi) = \lambda \varphi (\xi), \qquad \nu, \lambda, \, \in
{\mathbb C}.\ee  Proceedings as before and considering the results
of Appendix \ref{sec-appa}, the normalizable solutions of this
last equation, when $k_0= 1,2,3,$ are given respectively by the
deformed coherent symbols \be \varphi^{(1)}(\xi) = C_0 \exp\biggl(
(\lambda - \nu) \xi \biggr), \ee \be \varphi^{(2)}(\xi)= C_0
\left[ 1 + z (\lambda - \nu) \nu \xi \right] \exp\biggl( (\lambda
-\nu) \xi\biggr)\ee and \beqa \varphi^{(3)} (\xi)&=&  C_0
\biggl\{1 + z (\lambda - \nu) \nu \xi
 + z^2 \biggl[\left( \frac{{\lambda }^2\,\nu
 }{2} \nonumber +
  2\,\lambda \,{\nu }^2  -
  \frac{3\,{\nu }^3}{2} \right) \xi   \\ &+&
  \left(
  \frac{{\lambda }^2\,{\nu }^2 }{2}   -
  \lambda \,{\nu }^3 +
  \frac{{\nu }^4}{2}\right) \, {\xi }^2 \biggr]  \biggr\} \exp\biggl( (\lambda -\nu) \xi
  \biggr).
  \eeqa
Theses solutions can be  normalized and represent zero, first and
second order paragrassmann deformations, respectively, of coherent
states associated to the standard harmonic oscillator. For higher
values of $ k_0,$ we must proceed as in Appendix \ref{sec-appa}.

\subsection{Deformed algebra eigenstates for \mathversion{bold} ${\cal U}_{z,p}
(h(2))$} It is interesting to compute the AES associated to ${\cal
U}_{z,p} (h(2)), \ z,p \ne 0, $ and compare it with the ones
associated to ${\cal U}_{z,0} (h(2)).$ As we have noticed in
section \ref{sec-two}, these quantum algebras are isomorphic in
the sense that there is a nonlinear change of basis transforming
one to the other. In general, the existence of this isomorphism
does not imply the existence of an internal homomorphism  at the
AES level. Indeed, by definition, the eigenvalue equation
determining the set of AES deals with an arbitrary linear
combination of the deformed algebra generators, then from the
inverses of transformations \eqref{optilde} and  the solvable
structure of the commutation relations \eqref{com-he1tilde}, it is
impossible to find an internal homomorphism, at the AES level,
transforming the eigenvalue equation with $z,p \ \ne 0 $ to the
eigenvalue equation with $z \ne 0, p=0.$

To see that, in this section, we consider the two parameters
deformed algebra ${\cal U}_{z,p} (h(2))$ as given by
\eqref{com-he1}, and compute the AES using the particular
realization \eqref{op-def-one-zp}. More precisely, we have to
solve  the eigenvalue equation  \be \left[ e^{ z a^\dagger}
\sqrt{1+ {\left({p \over 2} e^{z a^\dagger}\right)}^2 } a + \mu
a^\dagger + {2 \nu \over p} \sinh^{-1} \left( {p\over 2} e^{z
a^\dagger} \right) \right] |\psi\rangle = \lambda |\psi\rangle,
\qquad \mu,\nu,\lambda \in {\mathbb C}. \ee In the Bargmann
representation, this equation becomes the first order differential
equation \be\left[ e^{ z \xi} \sqrt{1+ {\left({p \over 2} e^{z
\xi}\right)}^2 } {d\over d\xi} + \mu \xi + {2 \nu \over p}
\sinh^{-1} \left( {p\over 2} e^{z \xi} \right) \right] \psi(\xi) =
\lambda \psi (\xi), \qquad \mu,\nu,\lambda \in {\mathbb C}. \ee
When $z=0,$ we easily get the standard squeezed symbols \be
\psi_{o,p} (\xi) = C_0 (p,\lambda , \mu ,\nu)
\exp\left[\left(\lambda - {2\nu \over p} \sinh^{-1} (p/2)\right)
\xi - \mu {\xi^2 \over 2} \right]. \ee These symbols correspond to
the Bargmann representation of the AES associated to
 the deformed quantum Heisenberg algebra realization
\eqref{def1}. Moreover, when $p$ goes to zero, these symbols
becomes the standard squeezed symbols associated to $h(2).$

When $z \ne 0,$ making the change of variable $\zeta = e^{z\xi}, $
rearranging the terms and using the method of characteristics
curves to separate the differentials, we get \be {d\psi \over
\psi} (\zeta) = {\left[ \lambda - {\mu \over z} \ln \zeta - {2\nu
\over p} \sinh^{-1} {\left( p \zeta \over 2\right)} \right] \over
z \, \zeta^2\, \sqrt{1 + {p^2 \zeta^2 \over 4} }} d\zeta.\ee
Integrating both sides of this equation and then exponentiating,
we get \beqa \psi_{z,p} (\zeta) &=& C_0 (\lambda, \mu,\nu ; z, p )
\, \exp\Biggl[ {\sqrt{1 + {p^2 \zeta^2 \over 4}} \over z^2 \zeta}
\biggl( (1+ \ln \zeta) \mu - \lambda z + {2 \nu z \over p}
\sinh^{-1}(\frac{p \,\zeta}{2}) \biggr) \nonumber \\  &-& {\mu p
\over 2 z^2} \sinh^{-1}(\frac{p \,\zeta}{2}) - {\nu \over z} \ln
\zeta  \Biggr]. \label{so--gen-zp}\eeqa This result includes the
ones obtained for \eqref{eigen-10} when $p$ goes to zero.
Moreover, when we set also $\nu=0,$ we regain
\eqref{solgen-varphi2}.

\subsubsection{Perturbed two parameters deformation coherent and
squeezed states} Up to first order of approximation in $z$ and
$p^2,$ the deformed symbol \eqref{so--gen-zp} writes  \beqa
\psi_{z,p} (\xi) &\approx&  \tilde C_0 (\lambda, \mu,\nu ; z, p )
\biggl[1 + z \left( {\mu \xi^3 \over 3} - {\lambda \xi^2 \over 2}
\right) \nonumber \\&+& { p^2 \over 4} \left( {\mu \xi^2 \over 4}
- \left({\lambda \over 2} - {\nu \over 3}  \right)\xi \right)
\biggr] \, \exp\left( (\lambda - \nu) \xi - {1\over 2} \mu \xi^2
\right).  \eeqa  In the case $\mu = \delta e^{i \phi},$ $\lambda =
\beta e^{i \theta}$ and $\nu = - \gamma e^{i \eta},$ where $\gamma
\ge 0,$ a normalized version of these states, in the Fock
representation, is given by \beqa \nonumber |\psi \rangle &\approx
& {\tilde \Omega} (\delta, \phi, \beta, \theta, \gamma, \eta)
\biggl\{1 + \left[  z \left( {\delta e^{i \phi} \over 3}
{(a^\dagger)}^3 -  {\beta e^{i \theta} \over 2} {(a^\dagger)}^2
\right)  \right] \\ \nonumber &+& {p^2 \over 4} \left[{\delta e^{i
\phi} \over 4} {(a^\dagger)}^2 - \left({\beta e^{i \theta} \over
2} +  {\gamma e^{i \eta} \over 3} \right) a^\dagger
\right]\biggr\}
 \\ & & S \left( - \arctan (\delta )e^{i \phi} \right) D \left( {{\tilde
 \beta}
e^{i{\tilde \theta}} \over \sqrt{1 - \delta^2}} \right) |0
\rangle, \label{def-squee-sta}\eeqa where \beqa {\tilde \Omega}
(\delta, \phi, \beta, \theta, \gamma, \eta ) &=& 1  + {z  \over 2
{(1-\delta^2)}^2} \Biggl\{ {\tilde \beta} \Biggl[  \left( 2
\delta^2 +
{\tilde \beta}^2 \left( {1 + \delta^2 \over  1-\delta^2} \right)  \right)  \cos{\tilde \theta}  \nonumber \\
&-&  \delta  \left( 1 + \delta^2 +
 { 2 {\tilde \beta}^2 \over 1-\delta^2 } \right) \cos(\phi - {\tilde \theta})
\nonumber \\ &+& \delta^2  {\tilde \beta}^2 \left( 1  + {2
\delta^2 \over 3(1-\delta^2)} \right) \cos(2 \phi - 3 {\tilde
\theta}) - {2\delta {\tilde \beta}^2 \over 3(1-\delta^2)}
\cos(\phi - 3 {\tilde \theta}) \Biggr] \nonumber \\\nonumber &-&
\gamma \Biggl[ {\tilde \beta}^2 \cos(\eta - 2 {\tilde \theta}) -
\delta (2{\tilde \beta}^2 + 1 - \delta^2 )  \cos(\eta -  {\tilde
\theta}) \\ \nonumber &+& \delta^2  {\tilde \beta}^2 \cos(2 \phi-
\eta - 2 {\tilde \theta}) \Biggr]\Biggr\} - {p^2 \over 16
{(1-\delta^2)}^2} \Biggl\{ \delta {\tilde \beta}^2 (3 
\cos(\phi - 2 {\tilde \theta}) \\ &+&  {2\gamma\over 3}
 {\tilde \beta} (1 -\delta^2)   \biggl( \cos(\eta - {\tilde \theta}) + \delta \cos
(\phi -\eta - {\tilde \theta}\biggr) -
 2 {\tilde \beta}^2 - \delta^2 +
\delta^4  \Biggr\}, \nonumber \\
 \eeqa
where \be {\tilde \beta} = \sqrt{\beta^2 + \gamma^2 + 2 \beta
\gamma \cos(\eta-\theta)}, \qquad  {\tilde \theta}= \tan^{-1}
\left( { \beta \sin \theta + \gamma \sin \eta \over \beta \cos
\theta + \gamma \cos \eta } \right). \ee We notice that, in the
case $\gamma =0$ and $p=0,$ these normalized states become the
normalized  states given in equation \eqref{nor-z-def}.

\section{Some properties of the deformed  states}
\label{sec-cuatro} In this section, we will give some properties
of the deformed states found in preceding section. From Fock space
representation, we will deduce the physical quantities $X$ and
$P,$ representing the position and linear momentum of a particle,
respectively, and compute the corresponding dispersions in both
the perturbed deformed states associated to ${\cal U}_{z,p}
(h(2))$ and the deformed states associated to $ {\tilde {\cal
U}}_{z,0} (h(2)).$ We will also connect the last states with an
$\eta$-pseudo Hermitian Halmiltonian \cite{kn:AMostafazadeh}.

\subsection{Squeezing properties} \label{sec-xp-squeezed} First,
let us  consider the squeezing properties of $X$ and $P.$ In the
Fock space representation, these quantities are given by the
hermitian operators (we have assumed that the mass, angular
frequency and Planck's constant are all equal to 1) \be X = {(a +
a^\dagger) \over \sqrt{2}}, \qquad P= i {(a^\dagger - a )\over
\sqrt{2}}. \label{xp-def} \ee They verify the canonical
commutation relation \be [X,P] = i I . \ee The dispersion of these
quantities, computed on a specific normalized particle state
$|\psi\rangle,$ is defined as \be {(\Delta X )}^2 = \langle \psi |
X^2 | \psi \rangle - {(\langle \psi | X | \psi \rangle)}^2
\label{x-disper} \ee and \be {(\Delta P )}^2 = \langle \psi | P^2
| \psi \rangle - {(\langle \psi | P | \psi \rangle)}^2.
\label{p-disper} \ee The product of these dispersions satisfies
the Schr\"{o}dinger-Robertson uncertainty relation (SRUR)
\cite{kn:SchRo,kn:Mer} \be {(\Delta X )}^2 \, {(\Delta P )}^2 \ge
{1 \over 4} \biggl( \langle I \rangle^2 + \langle F \rangle^2
\biggr) =  {1 \over 4} \biggl( 1 + \langle F \rangle^2 \biggr) ,
\label{SchRo-prin} \ee where $F$ is the anti-commutator $ F = \{ X
- \langle X \rangle I, P- \langle P \rangle I\}.$ The mean value
of $F$ is a correlation measure between $X$ and $P.$ When $\langle
F \rangle = 0 ,$ we regain the standard Heisenberg uncertainty
principle.

 The minimum uncertainty states (MUS) are states that satisfy the
equality in \eqref{SchRo-prin}. They are called coherent states
when  the dispersions of both $X$ and $P$ are the same and
squeezed states when these  dispersions are different to each
other. The states for which the dispersion of $X$ is greater than
the one of $P$ are called $X$-squeezed whereas the states for
which the dispersion of $P$ is greater than the one of $X$ are
called $P$-squeezed.

We are interested to  compute the dispersions of $X$ and $P,$ in
the deformed squeezed states \eqref{def-squee-sta}, when $\nu=0,$
or $\gamma=0.$ More precisely, we want to study  the effect of the
deformation parameters on the squeezed properties of these
quantities.  As we have seen, when $z$ and $p$ go to zero, the
states \eqref{def-squee-sta} becomes the standard harmonic
oscillator squeezed states. In such a case, we know that the
dispersions of $X$ and $P$ are independent of $\lambda = \beta
e^{i \theta },$ and given by \cite{kn:NaVh1} \be {{(\Delta X )}_0
}^2 = {1 - 2 \delta \cos\phi + \delta^2 \over 2 (1-\delta^2)}
\qquad {\rm and} \qquad {{(\Delta P )}_0 }^2 = {1 + 2 \delta
\cos\phi + \delta^2 \over 2 (1-\delta^2)}. \ee  All these states
are MUS, that is, they satisfy the equality in \eqref{SchRo-prin}.

When $\gamma=0,$ the square of the mean value of $X,$ in the
states \eqref{def-squee-sta}, to first order of approximation in
$z$ and $p^2,$  is given by  \beqa \langle \psi |X |\psi \rangle^2
& \approx & 2 \biggl(\Remot \Gamma_{01}\biggr)  \, \Remot \Biggl\{
\biggl( 1+ 4 \epsilon (z,p) \biggr)  \Gamma_{01} \nonumber
\\\nonumber  &+& 2 z \, \Biggl( {\delta e^{-i \phi} \over 3}
\Gamma_{04} - {\beta e^{-i \theta} \over 2} \Gamma_{03} + {\delta
e^{i \phi} \over 3} \Lambda_{13} - {\beta e^{i \theta} \over 2 }
\Lambda_{12} \Biggr)
\\ &+& {p^2 \over 2} \, \Biggl( {\delta e^{-i \phi} \over
4} \Gamma_{03} - {\beta e^{-i \theta} \over 2} \Gamma_{02} +
{\delta e^{i \phi} \over 4} \Lambda_{12} - {\beta e^{i \theta}
\over 2} \Lambda_{11} \Biggr) \Biggr\}, \label{moyx2} \eeqa where
$\epsilon (z,p)= {\tilde \Omega}(\delta,\phi,\beta,\theta,0,0) - 1
$ and $\Gamma_{kl}$ and $\Lambda_{kl},$ $k,l=1,2,\ldots,$ are
matrix elements defined in Appendix \ref{sec-appb}. According to
\eqref{xp-def}, we have the same expression for the square of the
mean value of $P,$ but taking the imaginary part in place of the
real part.

On the other hand, the mean value of $X^2$ in the states
\eqref{def-squee-sta}, to first order of approximation in $z$ and
$p^2,$  is given by \beqa \nonumber \langle \psi |X^2 |\psi
\rangle & \approx & { 1 \over 2}+ \biggl( 1 + 2 \epsilon (z,p)
\biggr) (\Gamma_{11} +  \Remot  \Gamma_{02}) \\ \nonumber &+&  z
\, \Remot \Biggl( {\delta e^{-i \phi} \over 3} \Gamma_{05} -
{\beta e^{-i \theta} \over 2} \Gamma_{04} + {\delta e^{i \phi}
\over 3} \Lambda_{23} - {\beta e^{i \theta} \over 2 } \Lambda_{22}
\Biggr)
\\\nonumber &+& {p^2 \over 4} \, \Remot  \Biggl( {\delta e^{-i
\phi} \over 4} \Gamma_{04} - {\beta e^{-i \theta} \over 2}
\Gamma_{03} + {\delta e^{i \phi} \over 4} \Lambda_{22} - {\beta
e^{i \theta} \over 2} \Lambda_{21} \Biggr) \nonumber \\ & + &   z
\Biggl( {\delta e^{-i \phi} \over 3} (\Lambda_{41} - \Gamma_{03})
- {\beta e^{-i \theta} \over 2} (\Lambda_{31} - \Gamma_{02}) +
{\delta e^{i \phi} \over 3}(\Lambda_{14} - \Lambda_{03})\nonumber
\\ \nonumber &-& {\beta e^{i \theta} \over 2 } (\Lambda_{13} -
\Lambda_{02})\Biggr) + {p^2 \over 4} \Biggl( {\delta e^{-i \phi}
\over 4} ( \Lambda_{31} - \Gamma_{02})  - {\beta e^{-i \theta}
\over 2}  ( \Lambda_{21} - \Gamma_{01}) \nonumber \\&+& {\delta
e^{i \phi} \over 4} (\Lambda_{13} - \Lambda_{02}) - {\beta e^{i
\theta} \over 2} (\Lambda_{12} - \Lambda_{01}) \Biggr).
\label{x2moy} \eeqa Again, according to \eqref{xp-def}, we have
the same expression for the mean value of $P^2,$ but taking the
negative of the real part in place of the real part.

Combining \eqref{moyx2} with \eqref{x2moy}, according to equation
\eqref{x-disper}, we get the dispersion of $X$. In the same way,
we can obtain the dispersion of $P.$ Inserting the matrix elements
$\Gamma_{ij}$ and $\Lambda_{ij},$ as given in the Appendix
\ref{sec-appb}, we can compute these dispersions explicitly.

\begin{figure}[h] \centering
\begin{picture}(31.5,21)
\put(0,0){\framebox(31.5,21){}}
\put(1,2.1){\includegraphics[width=70mm]{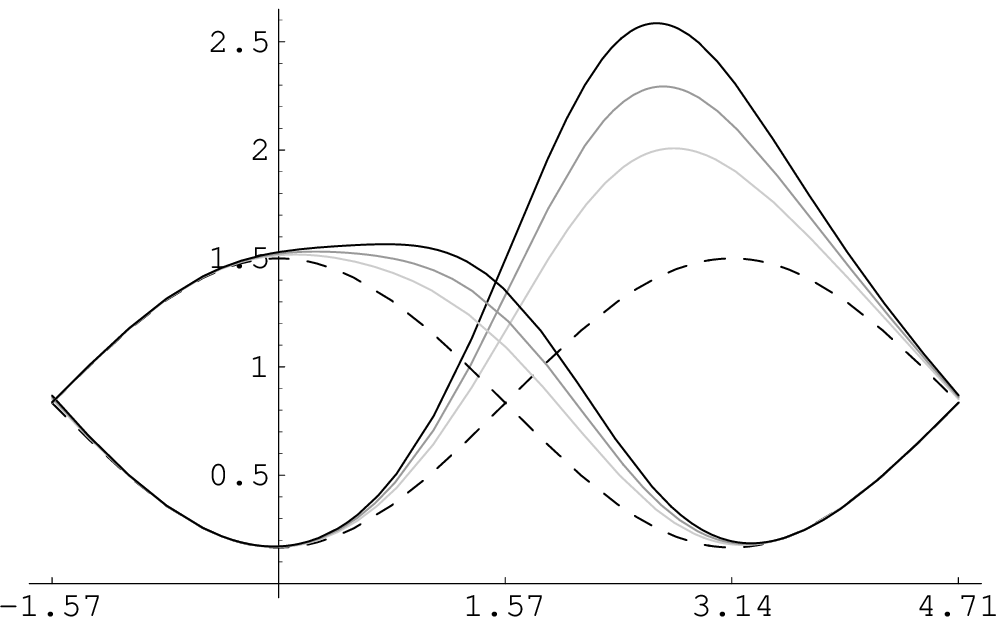}}
\put(28.9,3){\scriptsize{$\phi$}}
\put(27.6,5){\scriptsize{${(\Delta  P)}^2$}}
\put(26.9,5){\vector(-1,0){3}}
\put(25.6,15.4){\scriptsize{${(\Delta  X)}^2$}}
\put(25.4,15.4){\vector(-1,0){3}}
\end{picture}
\caption{Graphs of the dispersions of $X$ and  $P$ as functions of
$\phi$ for $p=0$ and $z=0.000,0.0010,0.0015,0.0020.$ }
\label{fig:varXPz0.0-0.020p=0}
\end{figure}

Figure \ref{fig:varXPz0.0-0.020p=0} show the dispersions of $X$
and $P$ in the minimum uncertainty squeezed states in dashed
lines, and in the deformed squeezed states in solid lines, as a
function of $\phi$ for fixed valued of the parameters $\delta,
\beta, \theta$ and $p$ ($\delta=0.5, \, \beta=2.0, \, \theta =0.8
\, \pi, p=0.00 $) and for special values of $z=0.0010, 0.0015,
0.0020$ (from the smaller to the greater gray level).
 We observe
that, as a consequence of the small deformations in the parameters
$z$ the squeezing properties of $X$ and $P$ have not been
essentially changed. Thus, in all the cases, we have $P$--squeezed
states when $- {\pi \over 2} < \phi < {\pi \over 2}, $ and
$X$--squeezed states when ${\pi \over 2} < \phi < {3 \pi \over 2}.
$ Also we observe that the product of the dispersions of $X$ and
$P$ in the deformed squeezed states, for a given value of $\phi,$
is always greater than the product of the dispersions in the
minimum uncertainty states, as required by the SRUR. These
difference is more remarkable  for values of $\phi$ in the range
${\pi \over 2} \le \phi <  {3\pi \over 2}. $ Let us notice that
when $\phi = \pm {\pi \over 2},$ the MUS are coherent states, in
the sense of the SRUR, i,e., the dispersion of  $X$ and $P,$ are
the same. Indeed, in all these cases, ${{(\Delta X)}_0}^2 =
{{(\Delta P)}_0}^2 = 0.83. $ This value is conserved by the
product of the dispersions of $X$ and $P$ in the deformed squeezed
states when $\phi = - {\pi \over 2},$ but  when $ \phi = {\pi
\over 2},$ it grows quickly as $z$ increases.

\begin{figure}[h]
\centering
\begin{picture}(31.5,21)
\put(0,0){\framebox(31.5,21){}}
\put(1,2.1){\includegraphics[width=70mm]{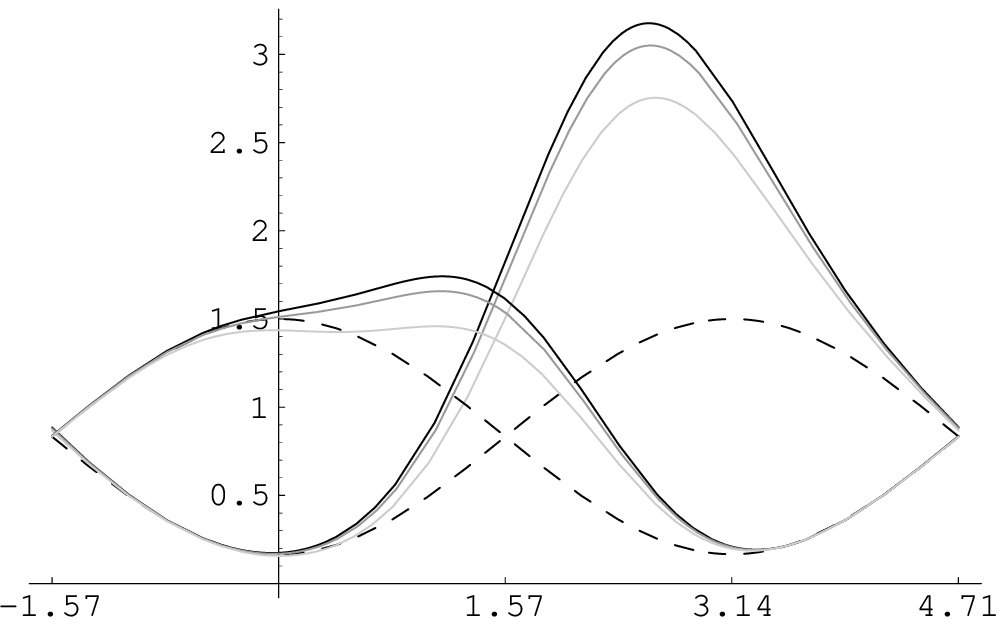}}
\put(28.9,3){\scriptsize{$\phi$}}
\put(27.8,5.5){\scriptsize{${(\Delta  P)}^2$}}
\put(27.5,5.5){\vector(-1,0){2.5}}
\put(25.3,15.4){\scriptsize{${(\Delta  X)}^2$}}
\put(25,15.4){\vector(-1,0){2.5}}
\end{picture}
\caption{Graphs of the dispersions of $X$ and  $P$ as functions of
$\phi$ for $z= 0.0030, p=0.00,0.06,0.11, \beta=2.0, \theta =0.8
\pi $ and $ \delta= {0.5}.$} \label{fig:varXPz-3p0-0.11}
\end{figure}

Figure \ref{fig:varXPz-3p0-0.11} show the dispersions of $X$ and
$P$ in the minimum uncertainty squeezed states in dashed lines,
and in the deformed squeezed states in solid lines, as a function
of $\phi$ for fixed valued of the parameters $\delta, \beta,
\theta$ and $z$ ($\delta=0.5, \, \beta=2.0, \, \theta =0.8 \, \pi,
z=0.0030 $) and for special values of $p=0.00, 0.06, 0.11$ (from
the greater to the smaller gray level).   We observe that the
product of dispersions of $X$ and $P$  decreases when $p$
increases. Thus the influence of the $p$ parameter on the first
order in $z$ deformed states is to reduce the uncertainty product
of $X$ and $P$ and to bring closer this quantity to the minimum
uncertainty values.
\begin{figure}[h]
\centering
\begin{picture}(31.5,21)
\put(0,0){\framebox(31.5,21){}}
\put(1,2.1){\includegraphics[width=70mm]{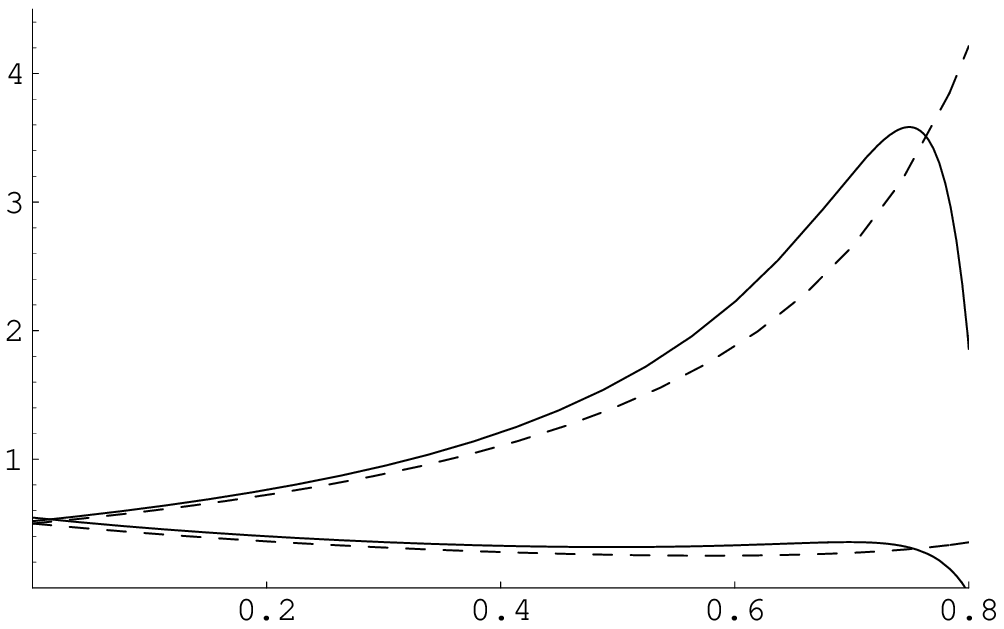}}
\put(28.9,3){\scriptsize{$\delta$}}
\put(27.6,7){\scriptsize{${(\Delta P)}^2$}}
\put(26.9,7){\vector(-1,-1){2}}
\put(14.9,13.4){\scriptsize{${(\Delta  X)}^2$}}
\put(16.8,12.9){\vector(1,-1){2}}
\end{picture}
\caption{Graphs of the dispersions of $X$ and  $P$ as functions of
$\delta$ for $z= 0.0025, p=0.01, \beta=2.0, \theta =0.8 \pi $ and
$ \phi= {\pi\over 6} .$} \label{fig:artvap01}
\end{figure}

Figure \ref{fig:artvap01} shows the typical behavior of the
dispersions of $X$ and $P$ in the minimum uncertainty squeezed
states in dashed lines, and in the deformed squeezed states in
solid lines, as a function of $\delta $ for $\phi=0.5, \,
\beta=2.0, \, \theta =0.8 \, \pi, z=0.0025 $ and $p=0.001.$ We
observe again that, as a consequence of the small deformations in
$z$ and $p,$ the squeezing properties of $X$ and $P$ have not been
essentially changed. Thus, the figure shows the behavior of
$P$--squeezed and $P$-deformed squeezed states. When $ 0< \delta
\lesssim 0.75,$ the product of the dispersions of $X$ and $P,$ in
the deformed squeezed states is always greater than the
corresponding product in the minimum uncertainty squeezed states,
as required by the SRUR.  For higher values of $\delta,$ only the
dashed lines represent the true behavior of the dispersions of $X$
and $P.$ Indeed, the approximation for the deformed squeezed
states, in this region, is not valid. These states are no longer
normalizable.
\subsection{General formulas for the dispersions of $X$ and $P$ in
the \boldmath $z$ deformed  states} The mean values  of $X^k,
k=1,2, \ldots,$ in the states \eqref{re-norm-squee} can be
expressed in the forme \be \biggl. \langle \varphi | X^{k} |
\varphi \rangle ={ {\partial^k \over
\partial \tau^k} \widetilde{\langle \varphi} | e^{\tau X} |
\widetilde{\varphi \rangle} \biggr|_{\tau=0} \over
\widetilde{\langle \varphi } | \widetilde{ \varphi \rangle}} ={
{\partial^k \over
\partial \tau^k}\biggl\{ e^{- {\tau^2 \over 4}} \ \widetilde{\langle \varphi} | e^{{\tau\over \sqrt2} a}
 e^{{\tau\over \sqrt2} a^\dagger}| \widetilde{\varphi \rangle} \biggr\}\biggr|_{\tau=0} \over
\widetilde{\langle \varphi } | \widetilde{ \varphi \rangle}}, \ee
where \be \widetilde{|\varphi \rangle} =   \exp\left( e^{-z
a^\dagger} {(\mu - \lambda z + \mu z a^\dagger) \over z^2} \right)
|0\rangle.  \ee Inserting these results into \eqref{x-disper} and
evaluating we get \be \label{x-disper-tau} {(\Delta X )}^2 = - { 1
\over 2} + { {\partial^2 \over
\partial \tau^2} \widetilde{\langle \varphi} | e^{{\tau\over \sqrt2} a}
 e^{{\tau\over \sqrt2} a^\dagger}| \widetilde{\varphi \rangle} \biggr|_{\tau=0} \over
\widetilde{\langle \varphi } | \widetilde{ \varphi \rangle}} -
{\Biggl( { {\partial \over
\partial \tau } \widetilde{\langle \varphi} | e^{{\tau\over \sqrt2} a}
 e^{{\tau\over \sqrt2} a^\dagger}| \widetilde{\varphi \rangle} \biggr|_{\tau=0} \over
\widetilde{\langle \varphi } | \widetilde{ \varphi \rangle}}
\Biggr)}^2 . \ee  To compute the  matrix element
$\widetilde{\langle \varphi} | e^{{\tau\over \sqrt2} a}
 e^{{\tau\over \sqrt2} a^\dagger} | \widetilde{\varphi \rangle}, $
we can firstly  write
 \be
 e^{{\tau\over \sqrt2} a^\dagger}| \widetilde{\varphi \rangle}
 = \sum_{n=0}^{\infty} C_n (\tau) |n\rangle
 \ee
and then to compute the coefficients $C_n (\tau), \ n=0,1,2,\ldots
, $ in the Bargman representation, in the same way  as we have do
it in section \eqref{sub-sec-coh-squee}. That is \be
\label{varphi-tilde-modif} \widetilde{\langle \varphi} |
e^{{\tau\over \sqrt2} a}
 e^{{\tau\over \sqrt2} a^\dagger} | \widetilde{\varphi \rangle} =
 \sum_{n=0}^{\infty} {\bar C}_{n} (\tau) C_{n} (\tau),
 \ee
where \be \label{cn-tau} C_n (\tau) = {1\over \sqrt{n!}}
\sum_{r=0}^{n}
 {n \choose r} {\left({\tau \over \sqrt2}\right)}^r   \ z^{n-r} \
\sum_{k=0}^{\infty} \sum_{m=0}^{{\tilde k}_< } {n-r \choose m}
{{(- k)}^{n-r-m} \over (k-m)!} {\left({\mu \over z^2 }\right)}^m \
{\left({\mu \over z^2} - {\lambda \over z}\right)}^{k-m}, \ee with
${\tilde k}_< $ the minimum between $k$ and $n-r .$

Inserting  \eqref{varphi-tilde-modif} into \eqref{x-disper-tau}
and evaluating again we get  \beqa \nonumber {(\Delta X )}^2 &=& -
{ 1 \over 2} + {\sum_{n=0}^\infty \Biggl. \biggl[ {\bar C}_n
(\tau) C_n^{\prime \prime} (\tau) + {\bar C}_n^{\prime \prime} C_n
(\tau) + 2 {\bar C}_n^{\prime} C_n^{\prime} (\tau) \biggr]
\Biggr|_{\tau=0} \over \sum_{n=0}^\infty {\bar C}_n (0) C_n(0) }
\\&-&  {\Biggl( {\sum_{n=0}^\infty \Biggl. \biggl[ {\bar C}_n
(\tau) C_n^{\prime} (\tau) + {\bar C}_n^{\prime} (\tau) C_n (\tau)
\biggr] \Biggl|_{\tau=0}\over \sum_{n=0}^\infty {\bar C}_n (0)
C_n(0) }\Biggr)}^2 ,\label{gen-for-x} \eeqa where, for instance,
$C_{n}^{\prime} (\tau) = {d C_n \over d\tau} (\tau) . $ From
\eqref{cn-tau}, we obtain \be C_{n}^{\prime} (0) ={1\over
\sqrt{n!}}
 {n \choose 1} {1 \over \sqrt2}   \ z^{n-1} \
\sum_{k=0}^{\infty} \sum_{m=0}^{{\tilde k}_1 } {n-1 \choose m}
{{(- k)}^{n-1-m} \over (k-m)!} {\left({\mu \over z^2 }\right)}^m \
{\left({\mu \over z^2} - {\lambda \over z}\right)}^{k-m}, \ee when
$n= 1, 2, \ldots, $ with ${\tilde k}_1 $ the minimum between $k$
and $n-1,$ \be C_{n}^{\prime \prime} (0) ={1\over \sqrt{n!}}
 {n \choose 2} \ z^{n-2} \
\sum_{k=0}^{\infty} \sum_{m=0}^{{\tilde k}_2 } {n-2 \choose m}
{{(- k)}^{n-2-m} \over (k-m)!} {\left({\mu \over z^2 }\right)}^m \
{\left({\mu \over z^2} - {\lambda \over z}\right)}^{k-m}, \ee when
$n= 2, 3, \ldots, $ with ${\tilde k}_2 $ the minimum between $k$
and $n-2,$ and  \be C_0^\prime (0) =  C_0^{\prime \prime} (0) =
C_1^{\prime \prime} (0)=0.\ee The formula to the dispersion of $P$
can be obtained from \eqref{gen-for-x} changing the $\tau $
argument of $C_n (\tau) $ by $ i \tau $ and then deriving and
evaluating to $\tau = 0.$ Thus,  dispersions formulas  of $X$ and
$P$ at all order in $z$ can be obtained. The first order
perturbation formulas of these dispersions  must correspond to the
dispersions obtained in the preceding subsection, in the limit
when $p$ goes to zero.

\subsection{\boldmath $\eta$-pseudo Hermitian and Hermitian Hamiltonians} In
this section we show that the subset of deformed coherent states
\eqref{eigen-10-aes}, corresponding to the eigenvalue $\lambda=0,$
are the coherent states associated to an $\eta$-pseudo Hermitian
Hamiltonian \cite{kn:AMostafazadeh} but also, up to a similarity
transformation, the coherent states associated to a Hermitian
Hamiltonian, both isospectral to the harmonic oscillator
Hamiltonian.  Indeed, when $\lambda=0,$ the eigenstates
\eqref{eigen-10-aes} correspond to the solutions of the eigenvalue
equation \be \label{eigen-10-reduc}  {\cal A} |\psi \rangle =  -
\nu  |\psi \rangle, \qquad \nu \in {\mathbb C}, \ee where ${\cal
A} =  a + \mu a^\dagger e^{- z a^\dagger}. $ These solutions can
be written in the form \be \label{eigenatilde} |\psi ; - \nu
\rangle = {\tilde N}_0 (\mu, - \nu, z )  \ G  (\mu , z ) \ e^{-
\nu a^\dagger} |0\rangle, \ee where \be G (\mu, z ) = \exp \left(
- \mu \ \sum_{k=0}^{\infty} {{(-z a^\dagger )}^k \over k!}
 {{(a^\dagger)}^2 \over (k+2)} \right) , \label{eigen-aes-reduc} \ee and $ {\tilde N}_0 \, (\mu, - \nu, z
)$ is a normalization constant.

Let us now to define the operator \be \label{H-pseudo} {\cal H} =
G \ a^\dagger a \ G^{-1}, \ee which  satisfies \be {\cal
H}^\dagger = \eta {\cal H} \eta^{-1},\ee where $\eta$ is the
hermitian operator \be \eta (\mu , z) = {(G^{-1})}^\dagger G^{-1}.
\ee Thus ${\cal H}$ is an $\eta$-pseudo Hermitian Hamiltonian
\cite{kn:AMostafazadeh}. Moreover, as \be G a^\dagger G^{-1} =
a^\dagger, \qquad G a G^{-1} = {\cal A},    \ee we get \be
\label{exp-H-pseudo} {\cal H} = a^\dagger {\cal A} = a^\dagger
\left( a + \mu a^\dagger e^{- z a^\dagger} \right) = a^\dagger a +
\mu e^{- z a^\dagger} {(a^\dagger)}^2 .\ee On the other hand, by
construction, it is easy to verify that \be\label{com-h2-hab} [
{\cal H}, {\cal A}]= - {\cal A}, \qquad [{\cal H}, a^\dagger]=
a^\dagger, \qquad [{\cal A}, a^\dagger]=1 \ee and  \be
\label{h-on-ezero}{\cal H} |E_0 \rangle = 0, \ee where \be |E_0
\rangle ={\tilde N}_0 \, (\mu, 0, z ) G (\mu, z) |0\rangle. \ee
This state is thus an  eigenstate of ${\cal A}$ corresponding to
the eigenvalue $\nu = 0.$ Thus, according to \eqref{com-h2-hab}
and \eqref{h-on-ezero}, the hamiltonian ${\cal H}$ is isospectral
to the harmonic oscillator Hamiltonian.  ${\cal A}$ represents an
annihilation operator for this system and their eigenstates
\eqref{eigenatilde} are the associated coherent states of ${\cal
H}.$

Let us mention that $\cal H$ verifies all the useful properties of
pseudo-Hermitian operators \cite{kn:Mostafazadeh-2004}. For
instance,  $\cal H$ is Hermitian on the physical Hilbert space
$\mathfrak{H}_{{\rm phys}}$ spanned by their corresponding
eigenstates $ | \psi_n \rangle \propto {(a^\dagger)}^n G
|0\rangle, \ n=0,1,2,\ldots,$ endowed with the positive-definite
inner product $\langle \cdot | \eta \ \cdot \rangle.$  Also, $\cal
H$ may be mapped to a Hermitian Hamiltonian ${\tilde {\cal H}}$ by
a similarity transformation ${\tilde {\cal H}}= {\hat \rho} {\cal
H} {\hat \rho}^{-1}, $ where ${\hat \rho}(\mu,z)=
\sqrt{\eta(\mu,z)}= \sqrt{{G^{-1}}^\dagger G^{-1}},$ is a
Hermitian operator on a Hilbert space $\mathfrak{H}$ formed of
same vectorial space $\mathfrak{H}_{{\rm phys}} $ but endowed with
the original inner product $\langle \cdot | \cdot \rangle.$ Thus,
in our case, according to \eqref{H-pseudo}, the Hermitian
Hamiltonian ${\tilde {\cal H}},$ is unitarily equivalent to the
standard harmonic oscillator Hamiltonian and is given by \be
\label{hami-tilde} {\tilde {\cal H}} =  {\hat \rho} \ G \
a^\dagger a \ G^{-1} \ {\hat \rho}^{-1}. \ee Indeed, \be {\hat
\rho} G {({\hat \rho} G)}^{\dagger} = {\hat \rho} G G^{\dagger}
{\hat \rho}^{\dagger} = {\hat \rho} \eta^{-1} {\hat \rho} = {\hat
\rho} {({\hat \rho}^2)}^{-1} {\hat \rho} = I \ee and \be  {({\hat
\rho} G)}^{\dagger} {\hat \rho} G = G^{\dagger} {\hat
\rho}^{\dagger} {\hat \rho} G =  G^{\dagger} {\hat \rho}^2 G =
G^{\dagger} {(G^{-1})}^{\dagger} G^{-1} G = I, \ee that is
${({\hat \rho} G)}^{\dagger}={({\hat \rho} G)}^{-1}, $ i.e., $
{\hat \rho} G $ is an unitary operator.

Let us notice that in absence of deformation ($z=0$) the operator
$\hat \rho $ is given by \be \label{ro-u-0}  \hat \rho (\mu,0) =
\sqrt{\exp\left({\bar \mu} {a^2 \over 2 }  \right)\exp\left( \mu {
{a^\dagger}^2 \over 2} \right)} = {\biggl[\exp\left( \int_{0}^{1}
[{\bar \mu} K_- + \mu K_+ + \varsigma (s) K_3 ] d s
\right)\biggr]}^{1\over2}, \ee where $K_- = {a^2 \over 2},$ $K_+ =
{{(a^\dagger)}^2 \over 2} $ and $K_3 = {1\over 4}(a a^\dagger +
a^\dagger a) $ are the standard bosonic realizations of the
$su(1,1)$ Lie algebra generators verifying  the commutation
relations \be [K_-, K_+] = 2 K_3, \qquad [K_3, K_{\pm}] = \pm
K_{\pm}\ee and \be \varsigma (s) = - 2 {d \over ds} \ln q(s) , \ee
where \be q(s)= \cosh(|\mu|(1-s)) + |\mu| \sinh(|\mu|(1-s)). \ee
 In this case, the Hamiltonian \eqref{hami-tilde}
becomes \be \label{htilde-ro-u-0} {\tilde {\cal H}} = {\hat \rho}
(\mu,0) \ [ a^\dagger a + \mu {(a^\dagger)}^2 ]\ {\hat \rho}^{-1}
(\mu,0), \ee and represents a Hermitian Hamiltonian describing two
photon processes in a single mode. To know the  explicit form of
this Hamiltonian we must firstly factorize  the operator
\eqref{ro-u-0} in the form of a product of exponential operators
of each  $su(1,1)$ generators and then insert  it into
\eqref{htilde-ro-u-0}. This process requires to solve  some
Ricatti type differential equations.

For small values of $z,$ the Hamiltonian \eqref{hami-tilde}
describes corrections to the energy of this system as a
consequence of the deformation. In general, when $z\ne 0,$ the
Hamiltonian \eqref{hami-tilde} represents multi-photon processes
in a single mode.

The generalized coherent states associated to the system described
by \eqref{hami-tilde}, can be easily obtained from the coherent
states associated to the standard harmonic oscillator. Indeed,
they are given by \be \label{gen-zmunu-ch} | \nu, z, \mu \rangle =
{\hat \rho} (\mu, z) \  G(\mu, z) D(\nu) |0\rangle,  \ee where
$D(\nu)$ is the standard unitary displacement operator defined at
the end of subsection \ref{sec-perturba-z-real}. These coherent
states correspond to the coherent states associated to the
pseudo-Hermitian Hamiltonian \eqref{exp-H-pseudo}, up to the
transformation $ {\hat \rho} (\mu, z),$ and are eigenstates of the
annihilation operator ${\tilde {\cal A}} = {\hat \rho} (\mu, z)\
{\cal A} \ {\hat \rho}^{-1} (\mu, z)$ corresponding to the
eigenvalue $\nu.$

\section{Conclusions}
In this paper, we have found some realizations of the deformed
quantum Heisenberg Lie algebra ${\cal U}_{z,p} (h(2)),$ in terms of
the usual creation  and annihilation operators associated with Fock
space representation of the standard harmonic oscillator. The method
used to get these realizations can be easily applied to find the
realizations of other quantum Hopf algebras and super-algebras, such
as the bosonic and fermionic oscillators Hopf algebras\cite{HLR96}
or the quantum super-Heisenberg algebra, that can also be obtained
by using the $R$-matrix approach.

We have computed the AES associated to ${\cal U}_{z,p} (h(2)).$ We
have seen that the set of AES contains the set of coherent and
squeezed states associated to the standard harmonic oscillator
system but also a new class of deformed coherent and squeezed
states, parametrized by the deformation parameters.  We have
studied the behavior of the dispersions of the position and linear
momentum operators of a particle in a class of perturbed squeezed
states and we have compared them with the behavior of these
dispersions in the minimum uncertainty squeezed states. Also we
have computed these dispersions on the deformed states associated
to ${\cal U}_{z,0} (h(2)),$ for all values of the $z$ parameter.
To first order in $z,$ these last dispersions reduce to the
perturbed ones obtained to ${\cal U}_{z,p} (h(2)),$ when p goes to
zero. Besides,  we have constructed a $\eta$-pseudo Hermitian
Hamiltonian \cite{kn:AMostafazadeh} to which a subset of the set
of algebra eigenstates associated to  ${\cal U}_{z,0} (h(2)),$ are
the coherent states.  From  this point of view, our deformed
states are linked to Hamiltonians presenting important physical
aspects \cite{kn:Mostafazadeh-2004}. Indeed, our pseudo-Hermitian
Hamiltonian verifies naturally all the  properties of
pseudo-Hermitian Hamiltonians such as the existence of associated
biorthonormal basis, resolution of the identity, positive-definite
inner product, physical Hilbert space, unitary and invertible
operators mapping the pseudo-Hermitian operators to the Hermitian
ones, etc. Thus, with the help of pseudo-Hermitian quantum
mechanics techniques we are allowed to compute, for instance, the
spectrum, the eigenstates and the associated coherent states of
complicated deformed Hermitian Hamiltonians describing
multi-photon processes in a single mode. Also, we can compute more
easily quantities such as mean values of physical observables  and
transition amplitudes. Moreover, it could be interesting to know,
at least for small values of the deformation parameter $z,$ the
explicit form of the resolution of the identity verified by the
generalized coherent states \eqref{gen-zmunu-ch}. Indeed, this
fact could have important consequences, for instance, in the study
of corrections to the time evolution of the quantum fluctuations
associated to the quadratures of the position and linear momentum
of a system characterized by a Hamiltonian describing  one and two
photon processes in a single mode \cite{kn:Wei-Min}. This is a no
trivial problem and it could be developed elsewhere.

On the other hand, we have found  new  classes of deformed
squeezed states, parametrized by a real paragrassmann number,
i.e., a number $z$ such that $z^{k_0}=0, $ for some $k_0 \, \in
{\mathbb N}. $ These states can be normalized, even if $z$ is
considered as a complex paragrassmann number. In this last case,
when $k_0 =2,$ we can should interpret $z$ as an odd complex
Grassmann number and compare this new  classes of deformed
squeezed states with the ones  associated to the
$\eta$-super-pseudo-Hermitian Hamiltonians \cite{kn:NaVh3}.

\section*{Acknowledgments} The author would like to thank V. Hussin for
valuable discussions and suggestions. He also thanks the referees
for valuable  suggestions about this article. The author's research
was partially supported by research grants from NSERC of Canada.

\renewcommand{\theequation}{\thesection.\arabic{equation}}

\appendix
\setcounter{equation}{0}\section{Solving a paragrassmann valued
differential equation}\label{sec-appa} In this appendix we are
interested  to solve the differential equation \be
\label{equa-para-z} \left[ {d \over d\xi} + (\mu \xi + \nu)
\sum_{l=0}^{k_0 -1} {(- z)^l \over l!} {d^l \over d \xi^l} \right]
\varphi (\xi) = \lambda \varphi (\xi), \qquad \mu,\nu,\lambda \, \in
{\mathbb C}, \ee where $k_0 \in {\mathbb N}, \, k_0 \ge 1,$ and $z$
is a paragrassmann generator such that $z^{k} =0, \, \forall k \ge
k_0. $

Let us assume a solution of the type \be \label{sol-type} \varphi
(\xi)= \sum_{k=0}^{k_0 -1} z^k \varphi_k (\xi). \ee Inserting this
solution into \eqref{equa-para-z}, we get \be
\left[\sum_{k=0}^{k_0 -1} z^k   {d \varphi_k
  \over d\xi} + (\mu \xi + \nu) \sum_{l=0}^{k_0 -1}
\sum_{k=0}^{k_0 -1} {{(-1)}^{l} (z)^{k+l} \over l!} {d^l \varphi_k
\over d \xi^l} \right]   = \lambda \sum_{k=0}^{k_0 -1} z^k
\varphi_k .\ee Identifying the coefficients of independent powers
$z^k, k=0,1,2,\ldots, k_0 -1,$ in this equality, we get the
following system of differential equations ($k=1, \ldots, k_0 -1$)
\beqa {d \varphi_k \over d\xi} + (\mu \xi + \nu) \sum_{l=1}^k
{{(-1)}^l \over l!}
  {d^l \varphi_{k-l} \over d\xi^l} &=&  \left[(\lambda -\nu) - \mu \xi
  \right] \varphi_k , \label{varphi-k}  \\ {d \varphi_0
\over d\xi} &=&  \left[(\lambda -\nu) - \mu \xi
  \right] \varphi_0. \label{varphi-0}
   \eeqa Let us  notice that we can solve this system of differential equations
proceeding by iteration. Indeed, from equation \eqref{varphi-0},
we get \be \varphi_0 (\xi) = C_0  \exp\left( (\lambda - \nu) \xi -
{1\over 2} \mu \xi^2 \right), \ee where $C_0$ is an arbitrary
integration constant.  Also, from equation \eqref{varphi-k},  for
a given value of $k,$ the general solution ${\varphi}_k (\xi)$ is
of the type \be \varphi_k (\xi) = \left[C_k + A_k (\xi) \right]
\exp\left( (\lambda - \nu) \xi - {1\over 2} \mu \xi^2 \right),
\qquad k=1,\ldots, k_0 - 1, \label{varphik}  \ee where the $C_k $
are arbitrary integration constants and $A_k (\xi )$ are functions
of $\xi$ which can be  determined by solving the system of
differential equations $(k=1,2,\ldots, k_0 -1 ) $ \beqa \nonumber
{d A_k \over d \xi} &= &  \exp\left( {1\over 2} \mu \xi^2 -
(\lambda - \nu) \xi \right) \\ & & (\mu \xi + \nu) \sum_{l=1}^k
{{(-1)}^{l+1} \over l!} {d^l   \over d\xi^l} \left[ ( C_{k-l} +
A_{k-l} ) \exp\left( (\lambda - \nu) \xi - {1\over 2} \mu \xi^2
\right) \right]. \label{akes}  \eeqa Using the Leibnitz's
derivation rule it is easy to prove that \beqa \nonumber
 \exp\left( {1\over 2} \mu \xi^2 - (\lambda - \nu) \xi\right) {d^l \over d\xi^l} \,    \left[ ( C_{k-l} + A_{k-l} ) \exp\left(
(\lambda - \nu) \xi - {1\over 2} \mu \xi^2 \right) \right]  = \\
 \sum_{m=0}^{l} {l \choose m} \Biggl[ C_{k-l} {(\lambda - \nu)}^{l-m}
{\left({\mu \over 2}\right)}^{m /2}  {(-1)}^m H_m \left(
\sqrt{{\mu\over 2 }} \xi\right) \nonumber \\ + {d^{l-m} A_{k-l}
\over d \xi^{l-m}} \sum_{s=0}^{m} {m \choose s} {(\lambda -
\nu)}^{m-s}{\left({\mu \over 2}\right)}^{s /2}  {(-1)}^s H_s
\left( \sqrt{{\mu\over 2 }} \xi\right)\Biggr], \eeqa where \be H_m
(x) = e^{x^2} {d^m  \over dx^m} e^{- x^2}, \qquad m=0,1, \ldots,
\ee are the Hermite polynomials.

Inserting these results into \eqref{akes} and integrating with
respect to  $\xi,$ we get \beqa \nonumber A_k (\xi) &=&
\sum_{l=1}^k \sum_{m=0}^{l}{{(-1)}^{l+1} \over l!} {l \choose m}
\int (\mu \xi + \nu) \biggl[ C_{k-l} {(\lambda - \nu)}^{l-m}
{\left({\mu \over 2}\right)}^{m /2} {(-1)}^m H_m \left(
\sqrt{{\mu\over 2 }} \xi\right) \\ &+& {d^{l-m} A_{k-l} \over d
\xi^{l-m}} \sum_{s=0}^{m} {m \choose s} {(\lambda -
\nu)}^{m-s}{\left({\mu \over 2}\right)}^{s /2} {(-1)}^s H_s \left(
\sqrt{{\mu\over 2 }} \xi\right)\biggr] d\xi, \label{akgen} \eeqa
when  $k=1,2,\ldots, k_0 -1. $ This system of integral equations
can be solved by iteration using the initial condition $ A_0 (\xi)
= 0. $ For instance, when $k_0 \ge 2, $ from equation
\eqref{akgen}, we get  \be A_1 (\xi) = \left[ (\lambda - \nu ) \nu
\xi + \mu ( \lambda - 2 \nu ) {\xi^2 \over 2} - \mu^2 {\xi^3 \over
3}\right] C_0. \ee When $k_0 \ge 3, $ from \eqref{akgen}, we get
\beqa A_2 (\xi) &=& \Biggl[ \left( \frac{{\lambda }^2\,\nu
 }{2} +
  \frac{\mu \,\nu   }{2} +
  2\,\lambda \,{\nu }^2  -
  \frac{3\,{\nu }^3}{2} \right) \xi \nonumber \\ &-&
  \left( \frac{{\lambda }^2\,\mu }{4} -
  \frac{{\mu }^2 }{4} -
  2\,\lambda \,\mu \,\nu -
  \frac{{\lambda }^2\,{\nu }^2 }{2} +
  \frac{9\,\mu \,{\nu }^2 }{4} +
  \lambda \,{\nu }^3 -
  \frac{{\nu }^4}{2}\right) \, {\xi }^2  \nonumber  \\ &+&
  \left( \frac{2\,\lambda \,{\mu }^2 }{3} +
  \frac{{\lambda }^2\,\mu \,\nu }{2} -
  \frac{3\,{\mu }^2\,\nu }{2} -
  \frac{3\,\lambda \,\mu \,{\nu }^2}
   {2} + \mu \,{\nu }^3 \right) \,{\xi }^3 \nonumber  \\ &+&
 \left( \frac{{\lambda }^2\,{\mu }^2 }{8} -
  \frac{3\,{\mu }^3 }{8} -
  \frac{5\,\lambda \,{\mu }^2\,\nu }
   {6} + \frac{5\,{\mu }^2\,{\nu }^2}
   {6}\right) \, {\xi }^4  \nonumber \\ &-& \left( \frac{\lambda \,{\mu }^3}{6}
   - \frac{{\mu }^3\,\nu }{3} \right) \,{\xi }^5 +
  \frac{{\mu }^4\,{\xi }^6}{18} \Biggr] \, C_0  \nonumber  \\ &+&
  \left( (\lambda -
\nu ) \nu \xi + \mu ( \lambda - 2 \nu ) {\xi^2 \over 2} - \mu^2
{\xi^3 \over 3}\right) C_1 . \eeqa

Finally, the general solution of the differential equation system
\eqref{equa-para-z}, is obtained by inserting \eqref{varphik} into
\eqref{sol-type}:   \be \varphi(\xi) = \left[\sum_{k=0}^{k_0 - 1}
z^k (C_k + A_k (\xi) )\right] \exp\left( (\lambda - \nu) \xi -
{1\over 2} \mu \xi^2 \right), \ee with $A_k (\xi)$ given in
equation \eqref{akgen}. We notice that there exists an independent
solution for each integration constant $C_k, \, k=0,1,\ldots, k_0
-1.$

In the case $\nu=0,$ equation \eqref{akgen} reduces to
($k=1,2,\ldots, k_0 -1$)\beqa \nonumber A_k (\xi) &=& \mu \,
\sum_{l=1}^k \sum_{m=0}^{l}{{(-1)}^{l+1} \over l!}  {l \choose m}
\int  \xi \,  \biggl[ C_{k-l} \, {\lambda }^{l-m} \, {\left({\mu
\over 2}\right)}^{m /2}  {(-1)}^m H_m \left( \sqrt{{\mu\over 2 }}
\xi\right) \\ &+& {d^{l-m} A_{k-l} \over d \xi^{l-m}}
\sum_{s=0}^{m} {m \choose s} {\lambda}^{m-s}{\left({\mu \over
2}\right)}^{s /2} {(-1)}^s H_s \left( \sqrt{{\mu\over 2 }}
\xi\right)\biggr] d\xi.  \label{agenuu}\eeqa

Thus, for instance, from equation \eqref{agenuu}, when $k_0 \ge 2,
$ we get \be A_1 (\xi) =  \mu \left( \lambda {\xi^2 \over 2} - \mu
{\xi^3 \over 3}\right) C_0. \ee When $k_0 \ge 3, $ we get \beqa
A_2 (\xi) &=& \Biggl[\left( \frac{{\mu }^2 }{4}  - \frac{{\lambda
}^2\,\mu }{4}
 \right) \, {\xi }^2   + \frac{2\,\lambda \,{\mu }^2 }{3} \, {\xi }^3 \nonumber
 + \left( \frac{{\lambda }^2\,{\mu }^2 }{8} -
  \frac{3\,{\mu }^3 }{8} \right) \, {\xi }^4  \nonumber - \frac{\lambda \,{\mu }^3}{6}
\,{\xi }^5  +
  \frac{{\mu }^4\,{\xi }^6}{18} \Biggr] \, C_0  \nonumber  \\ &+&
  \left( {\mu \lambda  \over 2} \xi^2  - {\mu^2 \over
3} \xi^3 \right) C_1 . \eeqa

In the case $\mu =0, $ equation \eqref{akgen} reduces to
($k=1,2,\ldots, k_0 -1$) \beqa \nonumber A_k (\xi) &=& \nu \,
\sum_{l=1}^k {{(-1)}^{l+1} \over l!}   \biggl[{(\lambda - \nu)}^l
\left( \xi C_{k-l} + \int A_{k-l} \, d\xi \right) \\ &+&
\sum_{m=0}^{l-1} {l \choose m} {(\lambda - \nu)}^m {d^{l-1-m}
 \over d \xi^{l-1-m}} \, A_{k-l} \biggr]  . \label{akgennu} \eeqa
For instance, from this last equation, when $k_0 \ge 2, $ we get
\be A_1 (\xi) = (\lambda - \nu) \nu \xi C_0 \ee and when
 $k_0 \ge 3, $ we get\be A_2 (\xi) = \Biggl[\left(
\frac{{\lambda }^2\,\nu
 }{2} +
  2\,\lambda \,{\nu }^2  -
  \frac{3\,{\nu }^3}{2} \right) \xi \nonumber  +
  \left(
  \frac{{\lambda }^2\,{\nu }^2 }{2}   -
  \lambda \,{\nu }^3 +
  \frac{{\nu }^4}{2}\right) \, {\xi }^2  \nonumber \Biggr] \, C_0  \nonumber
  +
  (\lambda -
\nu ) \nu \xi   C_1 . \ee

\setcounter{equation}{0}\section{Matrix elements} \label{sec-appb}
In section \ref{sec-xp-squeezed}, we need to compute the following
matrix elements: \be \Gamma_{kl} = \langle 0 | D^\dagger \left(
 {\beta e^{i \theta} \over \sqrt{1 - \delta^2}} \right) S^\dagger \left( -
\tan^{-1} (\delta )e^{i \phi} \right)  {a^\dagger}^k  a^l  S
\left( - \tan^{-1} (\delta )e^{i \phi} \right) D \left(
 {\beta e^{i\theta } \over \sqrt{1 - \delta^2}} \right)|0\rangle, \ee
and \be \Lambda_{kl} = \langle 0 | D^\dagger \left(
 {\beta e^{i \theta} \over \sqrt{1 - \delta^2}} \right) S^\dagger \left( -
\tan^{-1} (\delta )e^{i \phi} \right)  a^k {a^\dagger}^l  S \left(
- \tan^{-1} (\delta )e^{i \phi} \right) D \left(
 {\beta e^{i\theta } \over \sqrt{1 - \delta^2}} \right)|0\rangle, \ee
with $ k,l = 0,1,2, \ldots.$ Using the relation  \be S^\dagger
\left( - \tan^{-1} (\delta )e^{i \phi} \right) a S \left( -
\tan^{-1} (\delta )e^{i \phi} \right)= {1\over \sqrt{1 -\delta^2}}
\biggl( a - \delta e^{i \phi} a^\dagger\biggr), \ee we can write
them in the form\be \Gamma_{kl} = \langle 0 | D^\dagger \left(
 {\beta e^{i \theta} \over \sqrt{1 - \delta^2}} \right)   {{\biggl( a^\dagger - \delta e^{- i \phi} a \biggr)}^k {\biggl( a - \delta e^{i \phi} a^\dagger\biggr)}^l \over
 {(1-\delta^2)}^{{k+l \over 2}}} D \left(
 {\beta e^{i\theta } \over \sqrt{1 - \delta^2}} \right)|0\rangle
 \ee and \be \Lambda_{kl} = \langle 0 | D^\dagger \left(
 {\beta e^{i \theta} \over \sqrt{1 - \delta^2}} \right)   {{\biggl( a - \delta e^{i \phi} a^\dagger \biggr)}^k {\biggl( a^\dagger - \delta e^{-i \phi} a \biggr)}^l \over
 {(1-\delta^2)}^{{k+l \over 2}}} D \left(
 {\beta e^{i\theta } \over \sqrt{1 - \delta^2}} \right)|0\rangle, \ee
respectively. From the above expressions, it is clear  that \be
\Gamma_{0l} = {\bar \Gamma}_{l0} =\Lambda_{l0}={\bar
\Lambda}_{0l}, \quad  \Gamma_{ll} = {\bar \Gamma}_{ll}, \quad
\Lambda_{ll} = {\bar \Lambda}_{ll}, \qquad l=0,1, \ldots, \ee and
\be \Gamma_{kl} = {\bar \Gamma}_{lk}, \qquad \Lambda_{kl} = {\bar
\Lambda}_{lk}, \qquad k,l = 0,1, \ldots. \ee

We notice that the $\Gamma_{kl}$ matrix elements correspond to \be
{\partial^k \over
\partial \sigma^k} {\partial^l \over
\partial \tau^l} \langle 0 | D^\dagger \left(
 {\beta e^{i \theta} \over \sqrt{1 - \delta^2}} \right) {\exp\left[\sigma ( a^\dagger - \delta e^{-i
 \phi}a)\right]
 \over {(1-\delta^2)}^{k/2}} {\exp\left[\tau ( a - \delta e^{i
\phi}a^\dagger )\right] \over {(1-\delta^2)}^{l/2}}D \left(
 {\beta e^{i\theta } \over \sqrt{1 - \delta^2}} \right)|0\rangle,
\label{gama-fac} \ee when $\sigma$ and $\tau $ go to zero.
Applying the usual B.H.C. formula to disentangle the exponentials
factors, we get \beqa \exp\left[\sigma ( a^\dagger - \delta e^{-i
 \phi}a)\right]\exp\left[\tau ( a - \delta e^{i
\phi}a^\dagger )\right] =  \exp\left[\sigma \tau \delta^2 - {1
\over 2} \sigma^2 \delta e^{-i \phi } - {1 \over 2} \tau^2 \delta
e^{i \phi } \right]\nonumber \\ \exp\left[\left(\sigma  -\tau
\delta e^{i \phi}\right) a^\dagger\right] \exp\left[\left(\tau
-\sigma \delta e^{-i \phi}\right) a \right].  \eeqa Inserting this
result in \eqref{gama-fac}, and acting with the exponential
operators  on the coherent states, we get \beqa  \Gamma_{kl} &=&
{1 \over {(\sqrt{1-\delta^2})}^{k +l}} {\partial^k \over
\partial \sigma^k} {\partial^l \over
\partial \tau^l} \Biggl\{ \exp\left[\sigma \tau \delta^2 - {1 \over 2} \sigma^2 \delta e^{-i
\phi } - {1 \over 2} \tau^2 \delta e^{i \phi } \right]\nonumber \\
&& \exp\left[\left(\sigma  -\tau \delta e^{i \phi}\right) {\beta
e^{-i\theta} \over \sqrt{1-\delta^2}} \right] \exp\left[\left(\tau
-\sigma \delta e^{-i \phi}\right) {\beta e^{i\theta} \over
\sqrt{1-\delta^2}} \right]\Biggr\} {\Biggr|}_{\sigma=\tau=0} .
\eeqa The matrix elements $\Lambda_{kl}$ can be obtained in the
same way, we get
 \beqa  \Lambda_{kl} &=&
{1 \over {(\sqrt{1-\delta^2})}^{k +l}} {\partial^k \over
\partial \sigma^k} {\partial^l \over
\partial \tau^l} \Biggl\{ \exp\left[\sigma \tau  - {1 \over 2} \sigma^2 \delta e^{i
\phi } - {1 \over 2} \tau^2 \delta e^{-i \phi } \right]\nonumber \\
&& \exp\left[\left(\sigma  -\tau \delta e^{-i \phi}\right) {\beta
e^{i\theta} \over \sqrt{1-\delta^2}} \right] \exp\left[\left(\tau
- \sigma \delta e^{i \phi}\right) {\beta e^{-i\theta} \over
\sqrt{1-\delta^2}} \right]\Biggr\} {\Biggr|}_{\sigma=\tau=0} .
\eeqa For example, \beqa \Gamma_{00} &=& \Lambda_{00} =1, \qquad
\Gamma_{01} = {\bar \Lambda}_{01} = {\beta e^{i\theta} - \beta
\delta e^{i (\phi -\theta)} \over  (1-\delta^2)}, \nonumber
\\  \Gamma_{02} &=& {\bar \Lambda}_{02} = { \beta^2 e^{2i\theta} - \delta e^{i \phi} (2\beta^2 + 1 -
\delta^2 ) +  \beta^2 \delta^2  e^{2i (\phi -\theta)} \over
{(1-\delta^2)}^2} \nonumber \\
 \Gamma_{11} &=& \Lambda_{11} -1 = {\beta^2 ( 1+ \delta^2 ) + \delta^2 ( 1- \delta^2 ) - 2 \beta^2 \delta \cos(\phi-\delta) \over
 {(1-\delta^2 )}^2 }, \nonumber \\
\Lambda_{12} &=& \biggl[ \beta e^{-i \theta } \biggl( \beta^2 +
2\beta^2 \delta^2 + (2+\delta^2)(1-\delta^2)\biggr) - \beta \delta
e^{i( \theta
-\phi) } \biggl(2\beta^2 + \beta^2 \delta^2 + 3 (1-\delta^2)\biggr) \nonumber \\
&+& \beta^3 \delta^2 e^{i( 3\theta - 2 \phi) } - \beta^3 \delta
e^{i( \phi - 3 \theta)} \biggr] / {(1-\delta^2 )}^3 . \eeqa

\end{document}